\documentclass[twocolumn,times]{aastex62}
\usepackage{graphicx}
\usepackage[flushleft]{threeparttable}
\usepackage{blindtext}
\usepackage{amsmath}
\usepackage{mathtools}
\usepackage{multirow}
\usepackage{comment}
\usepackage{lineno}
\turnoffeditone
\newcommand{\lyxmathsym}[1]{\ifmmode\begingroup\def\b@ld{bold}
  \text{\ifx\math@version\b@ld\bfseries\fi#1}\endgroup\else#1\fi}

\begin{document}
\shortauthors{Park et al.}
\def\nar{New Astron.}
\def\na{New Astron.}

\title{\large \textbf{Scattering of Ly$\alpha$ Photons through the Reionizing Intergalactic Medium: I. Spectral Energy Distribution}}

\correspondingauthor{Hyunbae Park, Hyo Jeong Kim, and Kyungjin Ahn}
\email{hyunbae.park@lbl.gov; hyojeong@gist.ac.kr; kjahn@chosun.ac.kr}

\author[0000-0003-1187-4240]{Hyunbae Park}
\affil{Lawrence Berkeley National Laboratory, CA 94720-8139, USA}
\affil{Berkeley Center for Cosmological Physics, UC Berkeley, CA 94720, USA}
\affil{Kavli IPMU (WPI), UTIAS, The University of Tokyo, Kashiwa, Chiba 277-8583, Japan}

\author{Hyo Jeong Kim}
\affil{Department of Earth Sciences, Chosun University, Gwangju 61452, KOREA}
\affil{Gwangju Institute of Science and Technology, Gwangju 61005, South Korea}

\author[0000-0003-3974-1239]{Kyungjin Ahn}
\affil{Department of Earth Sciences, Chosun University, Gwangju 61452, KOREA}

\author[0000-0002-4362-4070]{Hyunmi Song}
\affil{Department of Astronomy and Space Science, Chungnam National University, Daejeon 34134, Republic of Korea}

\author[0000-0003-1187-4240]{Intae Jung}
\affil{Department of Physics, The Catholic University of America, Washington, DC 20064, USA}
\affil{Astrophysics Science Division, Goddard Space Flight Center, Greenbelt, MD 20771, USA}
\affil{Center for Research and Exploration in Space Science and Technology, NASA/GSFC, Greenbelt, MD 20771}

\author{Pierre Ocvirk}
\affil{Observatoire Astronomique de Strasbourg, 11 rue de l'Universite, 67000 Strasbourg, FRANCE}

\author[0000-0002-0410-3045]{Paul R. Shapiro}
\affil{Department of Astronomy, University Texas, Austin, TX 78712-1083, USA}

\author{Taha Dawoodbhoy}
\affil{Department of Astronomy, University Texas, Austin, TX 78712-1083, USA}

\author{Jenny G. Sorce}
\affil{Université Paris-Saclay, CNRS, Institut d'Astrophysique Spatiale, 91405, Orsay, France}
\affil{Leibniz-Institut f\"{u}r Astrophysik, An der Sternwarte 16, 14482 Potsdam, Germany}

\author[0000-0002-5174-1365]{Ilian T. Iliev}
\affil{Astronomy Centre, Department of Physics \& Astronomy, Pevensey III Building, University of Sussex, Falmer, Brighton, BN1 9QH, UK}

\begin{abstract}
During reionization, a fraction of galactic Ly$\alpha$ emission is scattered in the intergalactic medium (IGM) and appears as a diffuse light extending megaparsecs from the source. We investigate how to probe the properties of the early galaxies and their surrounding IGM using this scattered light. We create a Monte Carlo algorithm to track individual photons and reproduce several test cases from previous literature. Then, we run our code on the simulated IGM of the CoDaII simulation. We find that the scattered light can leave an observable imprint on the emergent spectrum if collected over several square arcminutes. Scattering can redden the emission by increasing the path lengths of photons, but it can also make the photons bluer by upscattering them according to the peculiar motion of the scatterer. The photons emitted on the far blue side of the resonance appear more extended in both frequency and space compared to those emitted near the resonance. This provides a discriminating feature for the blueward emission, which cannot be constrained from the unscattered light coming directly from the source. The ionization state of the IGM also affects the scattered light spectrum. When the source is in a small HII region, the emission goes through more scatterings in the surrounding HI region regardless of the initial frequency and ends up more redshifted and spatially extended. This can result in a weakening of the scattered light toward high $z$ during reionization. Our results provide a framework for interpreting the scattered light to be measured by high-$z$ integral-field-unit surveys. 
\end{abstract}

\section{Introduction} \label{sec:intro}
A substantial fraction of high-$z$ galaxies during reionization emit strongly in Ly$\alpha$ due to the recombination in the photoionized or collisionally ionized gas in star-forming regions \citep{2014PASA...31...40D}. Along with the Lyman break feature used to detect the Lyman break galaxies (LBGs), the Ly$\alpha$ emission is the main tool for identifying high-redshift galaxies. 

UV photons with energies above the Lyman limit (13.6 eV) are a main source of Ly$\alpha$ radiation (\citealt{Partridge1967}). While star-forming clumps in the interstellar medium are considered the major source of Ly$\alpha$, the ionized intergalactic medium (IGM) can also work as a diffuse Ly$\alpha$ source at large scales \citep[e.g.,][]{Fernandez2006}. Strong ionizing radiation from active galactic nuclei or highly star-forming galaxies can also ``illuminate" nearby cold non-star-forming clouds and turn them into Ly$\alpha$ emitters \citep[LAEs;][]{2012MNRAS.425.1992C,2012MNRAS.423..344R}.

High-$z$ LAEs are being considered as a promising probe of reionization for next-generation surveys. During reionization, the IGM would remain neutral in regions that are far from bright ionizing sources. In these HI regions, the Ly$\alpha$ emission would be suppressed due to the extended damping-wing cross section of a hydrogen atom. Indeed, a steep decline in the LAE number density is found above $z=6$, in contrast to its more gradual evolution at lower redshifts. This decline appears steeper for fainter galaxies ($M_{\rm UV}\gtrsim -20$) \citep{2004ApJ...617L...5M,Fontana2010,Ouchi2010,Pentericci2011,2011ApJ...728L...2S, 2012MNRAS.422.1425C,2012ApJ...744...83O,2012ApJ...760..128M,2013ApJ...775L..29T,2014ApJ...794....5T,2017ApJ...842L..22Z,2021MNRAS.502.6044E}, while the decline appears less dramatic for the brighter ones \citep{2015MNRAS.451..400M,2018ApJ...863L...3C,2019ApJ...877..146J,2019ApJ...883..142H,2019ApJ...879...28H,2020ApJ...904..144J,2020ApJ...891L..10T,2021NatAs...5..485H,2021arXiv211114863J}, consistent with theoretical expectations for the ``inside-out'' reionization scenario \citep[e.g.,][]{2004ApJ...617L...5M,2017ApJ...839...44S,2018ApJ...857L..11M,2019MNRAS.487.5902K,2021MNRAS.508.3697G,2021ApJ...922..263P,2022MNRAS.510.3858Q,2022MNRAS.512.3243S}.

Since the IGM has a negligible amount of dust in the high-$z$ universe, most of the scattered light would make it to the present-day universe in the form of extended diffuse light around the source galaxies. We expect nearly 100\% of the sky is covered by this kind of diffuse light \citep{2018Natur.562..229W}, as suggested by recent observations \citep{2017A&A...608A...8L}. Ly$\alpha$ blobs (LABs; e.g.,  \citealt{Francis1996,Steidel2000,Matsuda2004,Matsuda2012,2014ApJ...793..114Y,2020ApJ...894...33K}) are considered to be the scattered light characterized by high Ly$\alpha$ luminosity ($\sim10^{43}$\textendash $10^{44}~{\rm erg}~{\rm s}^{-1}$) and their spatial extent of $\sim$$30$\textendash 200 kpc although the emission may also be powered by collisional excitation and photoionization due to dynamical interactions between galaxies and the nearby IGM (e.g., \citealt{Haiman2000}; \citealt{Dijkstra2009}; \citealt{Faucher-Giguere2010}). Many LABs are in proximity to LAEs (\citealt{Matsuda2004}), but some are associated with LBGs \citep{Steidel2000} or active galactic nuclei (e.g., \citealt{Bunker2003}; \citealt{Colbert2011}).

The scattered light can extend to megaparsecs with low surface brightness \citep{2011ApJ...739...62Z}. \citet{2018MNRAS.481.1320C} reported a direct detection of the emission up to $15$ comoving Mpc (cMpc hereafter) around a QSO at $z\sim2-3.5$. Simulation studies find that such large-scale Ly$\alpha$ emission is spatially correlated with the location of LAEs and depends on the IGM ionization state and the intrinsic emission spectrum of the source \citep[e.g.,][]{2012MNRAS.424.2193J, 2018ApJ...863L...6V}. Recent narrowband surveys are starting to statistically detect the scattered light at $z\sim6$ as a cross-correlation signal between the LAEs and the surface brightness, although the reionization effect has not been confirmed yet \citep{2021ApJ...916...22K,2021arXiv210809288K}. 

Therefore, understanding the transfer process of Ly$\alpha$ radiation is essential for constraining reionization from scattered light. The randomness of the scattering process requires a Monte Carlo-type calculation to reproduce the observation. Theoretical studies have evolved from assuming a simplified (e.g., static and symmetric) configuration around a source \citep{1973MNRAS.162...43H,1990ApJ...350..216N,1999ApJ...524..527L, 2000JKAS...33...29A,2001ApJ...554..604A,2002ApJ...567..922A, 2002ApJ...578...33Z,2006ApJ...645..792T} to more
realistic configurations (nonstatic, three-dimensional, or both: \citealt{2002ApJ...567..922A,2002ApJ...578...33Z, 2005ApJ...628...61C, 2006ApJ...649...14D, 2006ApJ...645..792T, 2006A&A...460..397V, 2007A&A...474..365S, 2009ApJ...696..853L,2012MNRAS.424..884Y}). Dust can substantially reduce the Ly$\alpha$ escape fraction $f_{\alpha}$ \citep{2006MNRAS.367..979H,2006A&A...460..397V,2008A&A...491...89V,2008A&A...480..369S,Yajima2012,2014MNRAS.441.2861H}, but this effect is unimportant in the IGM, which is presumably dust-free. Due to the extreme dynamic range of the interstellar density field, it is still challenging to obtain converged results from modern high-resolution hydrodynamic simulations \citep{2021ApJ...916...39C}. However, the IGM-scale calculation is relatively free from this issue, owing to the much milder dynamic range of the IGM density.

Our primary goal is to understand how the properties of the scattered Ly$\alpha$ light depend on the ionization states of the IGM and the properties of the source galaxies so that we can constrain them from future observations. To this end, we (1) develop our own 3D Monte Carlo Ly$\alpha$ transfer code, and (2) simulate the Ly$\alpha$ photons' scattering process in the IGM during reionization. Our Ly$\alpha$ transfer code includes the essence of the already existing calculation schemes by, e.g., \citet{2002ApJ...578...33Z}, \citet{2005ApJ...628...61C}, \citet{2006ApJ...645..792T}, \citet{2006ApJ...649...14D}, \citet{2006A&A...460..397V}, \citet{2007A&A...474..365S}, \citet{2009ApJ...696..853L}, and \citet{Yajima2012}. On top of this, we shall make extra efforts to implement the cosmological redshift effect and to interpolate the discrete mesh quantities. Our work revisits some of the work by \citet{2010ApJ...716..574Z} with a higher-resolution data set and an inhomogeneous ionization field of the IGM.

The base field for a realistic Ly$\alpha$ radiative transfer (RT) calculation is given by the Cosmic Dawn II simulation \citep[CoDaII;][]{2020MNRAS.496.4087O}. CoDaII solves fully coupled radiation transfer, hydrodynamics, and gravity to reproduce the density/velocity/ionization/temperature fields during the reionization era in a cosmological volume of $[64h^{-1}~{\rm Mpc}]^3$ on a $4096^3$ mesh. CoDaII reproduces the observed statistical properties of galaxies at $z\gtrsim 6$ \citep{2016MNRAS.463.1462O} and fits into the current constraints on reionization \citep{2020MNRAS.496.4087O}. Its mesh-type output data make it is suitable for Monte Carlo Ly$\alpha$ RT calculation.

This paper is organized as follows. In Section 2, we describe the Monte Carlo Ly$\alpha$ transfer code and the relevant equations. In Section 3, we reproduce the known analytic solutions of test problems and validate the accuracy of our code. In Section 4, we present the results of applying our code to the CoDaII data. We summarize and discuss our results in Section 5.

\section{Method}
\label{sec:method}

\subsection{Basic Equations}
\label{subsec:basicEQ}

The distance that a photon propagates until being scattered is determined by the optical depth of its path. The optical depth to the scattering ($\tau_s$) is drawn from the exponential probability distribution of $P(\tau_s)=e^{-\tau_s}$. For a photon emitted at a frequency $\nu$ toward a direction $\bold{\hat{n}}_i$ from a location $\bold{r}$, the optical depth for a propagation distance $s$ can be calculated from the HI number density $n_{\rm HI}$, the gas temperature $T$, and the bulk gas velocity $\bold{V}_{\rm pe}$. Specifically, the thermal velocity of H atoms in the propagation direction, $v_{\parallel}\equiv \bold{v}_{\rm th} \cdot \bold{\hat{n}}_i$, and the bulk motion in the propagation direction, $V_\parallel\equiv \bold{V}_{\rm pe} \cdot \bold{\hat{n}}_i$, enter the equation:
\begin{equation} \label{eq:tau}
\tau_\nu(s) \equiv \int_0^s ds^\prime n_{\rm HI}(\bold{r}^\prime) \int_{-\infty}^{\infty}dv_{\parallel}\,P(v_{\parallel};\bold{r}^\prime)\,\sigma(\nu^\prime(\bold{r}^\prime)).
\end{equation}
Here, $\bold{r}^\prime=\bold{r}+s^\prime \bold{\hat{n}}_i$ is the photon location after the propagation, $P(v_\parallel)$ is the probability distribution of $v_\parallel$, and $\sigma(\nu^\prime)$ is the Ly$\alpha$ cross section as a function of the photon frequency in the H atom frame $\nu^\prime$. In the H atom frame, the frequency is shifted from the original value at the emission, $\nu$, according to the peculiar motion of the atom and the cosmological redshift during the propagation:
\begin{equation}
\nu^\prime(\bold{r}^\prime)=\nu-\nu \frac{v_\parallel+V_\parallel(\bold{r}^\prime)+s^\prime H(z)}{c},
\end{equation}
where $H(z)$ is the cosmological expansion rate and $c$ is the speed of light. For a gas temperature $T$, the thermal velocity distribution is given by 
\begin{equation}
P(v_{\parallel};\bold{r}^\prime)=\frac{1}{\sqrt{\pi} v_{\rm th}(\bold{r}^\prime)}\exp\left(-\frac{v_{\parallel}^{2}}{v_{\rm th}^{2}(\bold{r}^\prime)}\right),\label{eq:v_dist}
\end{equation}
where $v_{\rm th}(\bold{r}^\prime)=\sqrt{2k_{\rm B}T(\bold{r}^\prime)/m_{\rm H}}$ is the mean thermal velocity of hydrogen atoms at the location $\bold{r}^\prime$, $k_{\rm B}$ is the Boltzmann constant, and $m_{\rm H}$ is the mass of a hydrogen atom. The Ly$\alpha$ scattering cross section is given by
\begin{equation}
\sigma(\nu)=f_{12}\frac{\pi e^{2}}{m_{e}c}\frac{{\Delta\nu_{L}}/{2\pi}}{(\nu-\nu_{0})^{2}+({\Delta\nu_{L}}/{2})^{2}},\label{eq:cross_section}
\end{equation}
where $f_{12}=0.4167$ is the Ly$\alpha$ oscillator strength, $e$ is the electron charge, $m_e$ is the electron mass, $\Delta\nu_{L}=9.936\times10^{7}~{\rm Hz}$ is the natural line width, and $\nu_{0}=2.466\times10^{15}~{\rm Hz}$ is the Ly$\alpha$ frequency.

Plugging Equations~(\ref{eq:v_dist}) and (\ref{eq:cross_section}) into Equation~(\ref{eq:tau}) gives
\begin{eqnarray} 
\tau_{\nu}&=&34.61\left(\frac{T}{10^{4}~\rm K}\right)^{-0.5}\nonumber\\
&&\times\int_{0}^{s}\left(\frac{ds^\prime}{\rm kpc}\right)\,\left(\frac{n_{\rm H}(\bold{r}^\prime)}{1.899\times 10^{-7}{\rm cm}^{-3}}\right)\mathcal{H}(a,x).
\end{eqnarray}
Here, $\mathcal{H}$ is the Voigt function defined as 
\begin{equation}
\mathcal{H}(a,x)=\frac{a}{\pi}\int_{-\infty}^{\infty}\frac{e^{-y^{2}}}{(x-y)^{2}+a^{2}}\,dy,
\label{eq:voigt}
\end{equation}
where $a\equiv\Delta\nu_{_{L}}/2\Delta\nu_{_{D}}=4.702\times10^{-4}(T/10^4~{\rm K})^{-0.5}$ is the ratio of the natural line width to the Doppler line width $\Delta\nu_{D}=\nu_{0}(v_{\rm th}/c)$ and
\begin{equation}
x\equiv\frac{\nu-\nu_{0}(1+V_{\parallel}/c+Hs^\prime/c)}{\Delta \nu_{D}},
\end{equation}
is the dimensionless frequency in the gas frame.

In practice, the integral form of the Voigt function (Eq. \ref{eq:voigt}) becomes a nuisance in the numerical calculation, and therefore we instead use a fitting formula given by Equations~(7) and (8) of \citet{2006ApJ...645..792T}, which gives an error of less than 1\% for $T\gtrsim 2\,{\rm K}$.

When the optical depth of the photon path reaches $\tau_s$, the photon is scattered by an H atom in a new direction $\bold{\hat{n}}_f$. In this work, the new direction $\bold{\hat{n}}_f$ is randomly drawn assuming the scattered photon is isotropically distributed\footnote{We note that this is a simplification. In the literature, the Rayleigh scattering with the probability distribution of $P(\mu) \propto 1+\mu^2$ with $\mu\equiv \bold{\hat{n}}_i\cdot\bold{\hat{n}}_f$ is usually considered. Nevertheless, this simplified, isotropic scheme has passed all our test problems in Section~\ref{sec:code test}, indicating that the effect of anisotropic scattering is smeared out after multiple scattering events. Therefore, we adopt this isotropic-scattering scheme for our calculation.}.

During the scattering event, the scattering atom experiences a small recoil $\delta v\sim h\nu_0/(m_{\rm H}c)$ of the order of a few m s$^{-1}$ depending on the difference between the incoming and outgoing directions of the photon. In the rest frame of the scattering atom, the energy transfer from this recoil is $\sim m_{\rm H}(\delta v)^2$, which has a negligible impact on the energy of the scattered photon. In the global frame, however, the energy transfer is $\sim m_{\rm H}v_{\rm atom} \delta v$, where the atom velocity $v_{\rm atom}=|\mathbf{v}_{\rm atom}|$ is of the order of a few km/s and makes a significant change to the photon energy, which is described by 
\begin{equation}
x_f=x_i
-\frac{\bold{v}_{\rm atom} \cdot \bold{\hat{n}}_i}{v_{\rm th}}
+\frac{\bold{v}_{\rm atom} \cdot \bold{\hat{n}}_f}{v_{\rm th}}
+g(\bold{\hat{n}}_i\cdot\bold{\hat{n}}_f-1)
\label{eq:nu_change}
\end{equation}
\citep[see also Sec. 7.3 of][]{2017arXiv170403416D}.
Here, $g=h\nu_{0}/(m_{\rm H}cv_{\rm th})\approx 2.6\times10^{-4} (T/10^4~\mbox{K})^{-0.5}$ is the recoil factor in the atom frame, and the atom velocity $\mathbf{v}_{\rm atom}$ is the sum of the bulk velocity of the gas and the thermal motion of the atom: $\bold{v}_{\rm atom} = \bold{V}_{\rm pe}+\bold{v}_{\rm th}$. We find that excluding the recoil term does not affect the results as was argued by \citet{1971ApJ...168..575A}, although we include it in our calculation.

In order to evaluate Equation~(\ref{eq:nu_change}), we need the thermal motion of the scattering atom $\mathbf{v}_{\rm th}$. The dimensionless thermal velocity parallel to the incident photon direction, $u_{\parallel}\equiv {\mathbf v}_{\rm th}\cdot \bold{\hat{n}}_i/v_{\rm th}$, is drawn from the probability distribution function,
\begin{equation}\label{eq:vel-par}
f(u_{\parallel})=\frac{a}{\pi \mathcal{H}(a,x)}\frac{e^{-u_{\parallel}^{2}}}{(u_{\parallel}-x)^{2}+a^{2}},
\end{equation}
which accounts for the simultaneous weighting by the thermal motion and the scattering cross section. We write another velocity component perpendicular to $\bold{\hat{n}}_i$ as $u_{\perp}\equiv |\textbf{v}_{\rm th}\times \bold{\hat{n}}_i|/v_{\rm th}$ and draw from a Gaussian probability distribution:
\begin{equation}
f(u_{\perp})=\frac{1}{\sqrt{\pi}}e^{-u_{\perp}^{2}}.
\label{eq:vel-per}
\end{equation}
We generate a random azimuthal angle $\phi$ from the flat distribution between $0$ and $2\pi$ for the perpendicular component to obtain the three-dimensional thermal velocity $\bold{v}^\prime_{\rm th}=v_{\rm th} (u_\perp\cos{\phi},u_{\perp}\sin{\phi},u_\parallel)$. 
We then apply to this vector a 3D rotation that moves $\bold{\hat{n}}_i$ to the $z$ direction to obtain the thermal velocity in the global frame $\bold{v}_{\rm th}$ to be used in Equation~(\ref{eq:nu_change}).

We note that discreteness in the physical quantities is unavoidable for calculations with numerical simulation outputs. 
In particular, the discreteness in the velocity field can easily lead to inaccurate results because of the steep dependence of $\sigma(\nu)$ on $V_{\parallel}$. Thus, we perform a 3D linear interpolation based on the eight nearest cell centers that enclose the location to calculate the physical quantities mentioned above. We find that enforcing continuity in the velocity field in this way dramatically reduces numerical artifacts throughout our calculation.

\subsection{Monte Carlo Simulation of Ly$\alpha$ Scattering} \label{sec:MC}

Given the stochasticity of the Ly$\alpha$ resonant scattering process, we adopt the Monte Carlo method for our calculation. We generate individual photons and track their paths as they propagate in space. The major steps of the Ly$\alpha$ scattering simulation using the Monte Carlo method are described below. 

\begin{itemize}

\item[Step] 1: Read 3D gridded data of gas density ($\rho$), hydrogen ionization fraction ($x_{\rm HII}$), peculiar velocity ($\mathbf{V}_{\rm pe}$), and temperature ($T$) fields from reionization simulation output.

\item[Step] 2: Create a photon with the initial frequency $\nu_i$, position $\mathbf{r}_i$, and direction $\bold{\hat{n}}_i$. 

\item[Step] 3: Draw a random optical depth $\tau_s$ from the scattering probability distribution, $P(\tau_s)=e^{-\tau_s}$. 

\item[Step] 4: Accumulate $\tau$ according to the propagation distance $s$. We propagate 0.1\% of the grid size at a time to evaluate Equation~(\ref{eq:tau}) while keeping track of the HI density, gas velocity, and temperature on the way.

\item[Step] 5: When the scattering happens after the optical depth reaches $\tau_s$, we update the position vector to be $\mathbf{r}_f=\mathbf{r}_{i}+s\bold{\hat{n}}_{i}$.

\item[Step] 6: Draw thermal velocity components for the atom that scattered the photon, $u_{\parallel}$ and $u_{\perp}$, from Equations~(\ref{eq:vel-par}) and (\ref{eq:vel-per}).

\item[Step] 7: Draw a new direction vector $\bold{\hat{n}}_{f}$ from an isotropic probability distribution. 

\item[Step] 8: Calculate the new frequency $\nu_{f}$ and the new direction vector $\hat{k}_f$ of the scattered photon using $\mathbf{V}_{\rm pe}$ and $\mathbf{v}_{\rm th}$.

\item[Step] 9: Replace the initial frequency and the initial direction vector by their final quantities from previous steps. Repeat Steps 3$-$8 until the photon escapes the simulation box. 

\item[Step] 10: Sample the location ($\bold{r}_{\rm es}$), direction ($\hat{\bold{k}}_{\rm es}$), and frequency ($\nu_{\rm es}$) at the final scattering before escape.

\item[Step] 11: Repeat Steps 2$-$10 until accumulating enough photons to draw statistics (typically $10^6$). 

\end{itemize}

We calculate the Ly$\alpha$ scattering in the source comoving frame and do not use any particular code acceleration method such as the core-skipping acceleration scheme \citep[e.g.,][]{2000JKAS...33...29A,2007A&A...474..365S}. Due to the serial nature of the calculation, the code can easily be parallelized to multiple cores with shared memory. The computational cost depends sensitively on the optical depth of the system as well as other parameters such as simulation box size, mesh grid, etc. In the application to the CoDaII data presented in Section~\ref{sec:CoDaII}, the calculation for $10^6$ photons takes several minutes to an hour depending on how neutral the IGM is around the galaxy.

\subsection{Cosmological Redshift} \label{sec:cosmo_effect}

\begin{figure}
\begin{center}
\includegraphics[width=0.4\paperwidth]{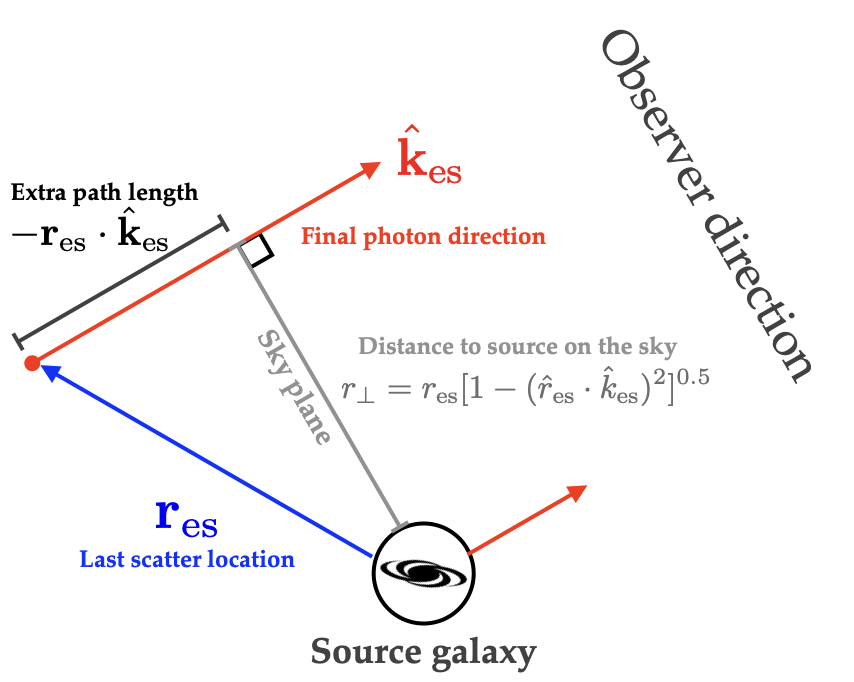}
\caption{\label{fig:ExtZ} Schematic description of the path length correction of Equation~(\ref{eq:vcorr}) for calculating the observed frequency $\nu_{\rm obs}$ from the final frequency at the last scattering event.}
\end{center}
\end{figure}

In our code validation tests presented in Section~\ref{sec:code test}, the physical sizes of the systems are small enough that cosmological redshift is negligible, and all the scattered photons are within the observational field of view. In this case, we can directly use the final frequency at the escape of the system ($\nu_{\rm es}$) to obtain the observed spectrum of the scattered light. 

For the application to the CoDaII simulation data in Section~\ref{sec:CoDaII}, however, the scattered light extends to several comoving megaparsecs and beyond, making cosmological redshifting an important factor in the spectrum of the scattered photons. Also, some of the scattered light may not be sampled depending on the transverse distance between the photon and the source, $r_\perp$, and the survey design. 

In order to calculate the observed spectrum, one must align the photons on the same sky plane with the source. In order to do so, we apply a path length correction to the final frequency of the last-scattered photons ($\nu_{\rm es}$) to obtain the frequency on the sky plane of the source galaxy ($\nu_{\rm obs}$):
\begin{equation}\label{eq:vcorr}
\nu_{\rm obs} = \nu_{\rm es} + \nu_{\rm es} c^{-1} H\mathbf{r}_{\rm es}\cdot\hat{\mathbf{k}}_{\rm es},
\end{equation}
where $\bold{r}_{\rm es}$ is the location of last scattering and $\hat{\bold{k}}_{\rm es}$ is the final photon direction. 
Then, the transverse distance to the source on the sky plane is given by
\begin{equation}
r_\perp=r_{\rm es} \sqrt{1-(\hat{\mathbf{r}}_{\rm es}\cdot\hat{\mathbf{k}}_{\rm es} )^2}.
\end{equation}
The additional path length term, $\mathbf{r}_{\rm es}\cdot\hat{\mathbf{k}}_{\rm es}$, in Equation~(\ref{eq:vcorr}) and the projected distance are illustrated in Figure~\ref{fig:ExtZ}.
We calculate $\nu_{\rm obs}$ and $r_\perp$ for each sampled photon, assuming the observer is in the final photon direction. We then combine the statistics of all the photons escaping in different directions, effectively averaging the observations of one halo from all sightlines.

\section{Code Validation Test} \label{sec:code test}

We test the Ly$\alpha$ scattering code for several simplified cases studied by previous works. 
We create mesh-type initial conditions for those models and run our Monte Carlo Ly$\alpha$ scattering code. 
We also reproduce some of the results in Section~\ref{sec:CoDaII} with another well-tested Ly$\alpha$ scattering code by \citet{2022ApJS..259....3S} and find a good agreement.
In particular, the first scattering location (defined in Sec.~\ref{sec:CoDaII}) is perfectly reproduced by their code because it does not involve any randomness.

\subsection{Static homogeneous slab}

\begin{figure}
\includegraphics[width=0.4\paperwidth]{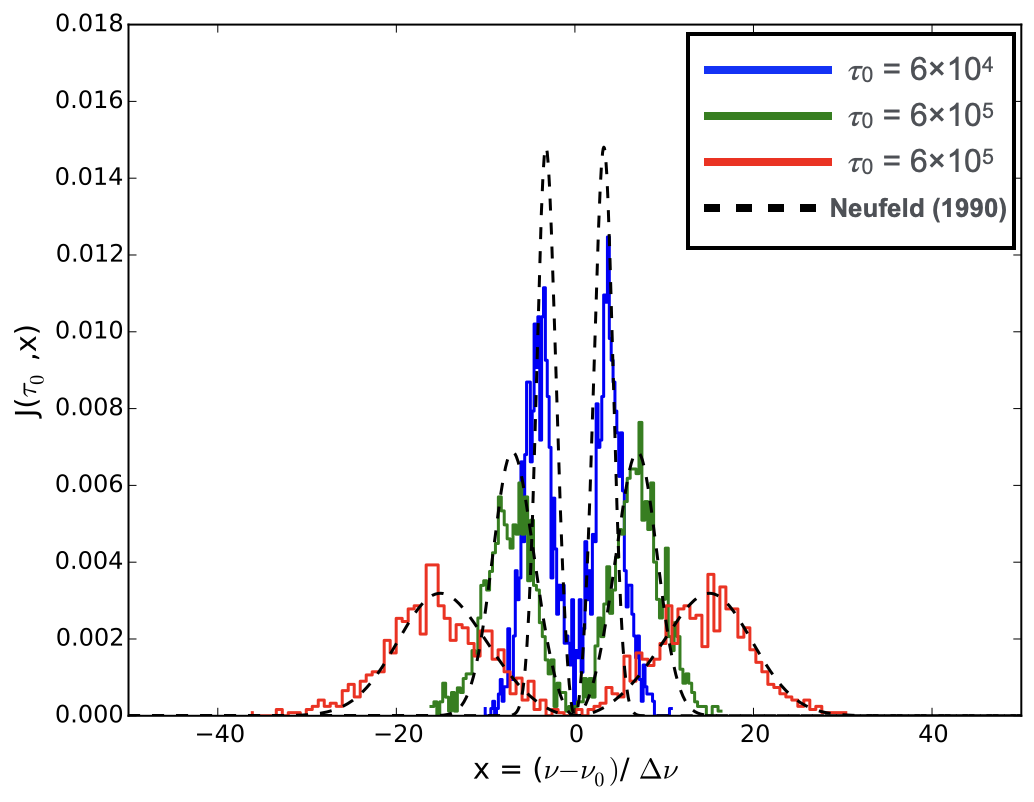}
\caption{\label{fig:Static-slab} Intensity of scattered Ly$\alpha$ photons from the static slab model. The agreement between the simulation (solid lines) and the analytic solution (dashed lines) improves as we increase the optical depth of the slab (blue$\rightarrow$green$\rightarrow$red) because the analytic solution was derived from the optically thick limit.}
\end{figure}

In the static homogeneous slab test, we locate a monochromatic source at the center of the slab. The slab is static and homogeneous with the neutral hydrogen column density $N_{\rm HI}$. We try $N_{\rm HI}=10^{18},10^{19},$ and $10^{20}\;{\rm cm}^{-2}$, which correspond to $\tau_{0}=$ $6\times10^{4}$, $6\times10^{5}$, and $6\times10^{6}$, respectively. We generate photons at the line center (i.e., $x=0$), and the line center optical depth $\tau_{0}$ is calculated from the slab center to the edge. The temperature of the slab is set to $10^4$ K everywhere. The photon escaping from the slab is collected to obtain the emergent spectrum.
\citet{1990ApJ...350..216N} solved the radiative transfer equation for this configuration and obtained an angular mean intensity of
\begin{equation}
J(\tau_0, x) =\frac{\sqrt{6}}{24}\frac{x^{2}}{\sqrt{\pi}a\tau_{0}}\frac{1}{\cosh[\sqrt{{\pi^{3}}/{54}}(x^{3}-x_i^{3})/a\tau_{0}]}.\label{eq:J-slab}
\end{equation}

We plot the emergent spectrum in Figure \ref{fig:Static-slab}. The solid lines are results from the Ly$\alpha$ code, and the dashed lines are the analytic solution from Equation (\ref{eq:J-slab}). The simulation reproduces the emergent spectrum of the analytic model very well. In the homogeneous slab case, photons diffuse symmetrically in the frequency domain and show the double-peak features. Equation (\ref{eq:J-slab}) was derived for the optically thick case, so the emergent spectrum fits better as $\tau_{0}$ increases.

\begin{figure}
\includegraphics[width=0.4\paperwidth]{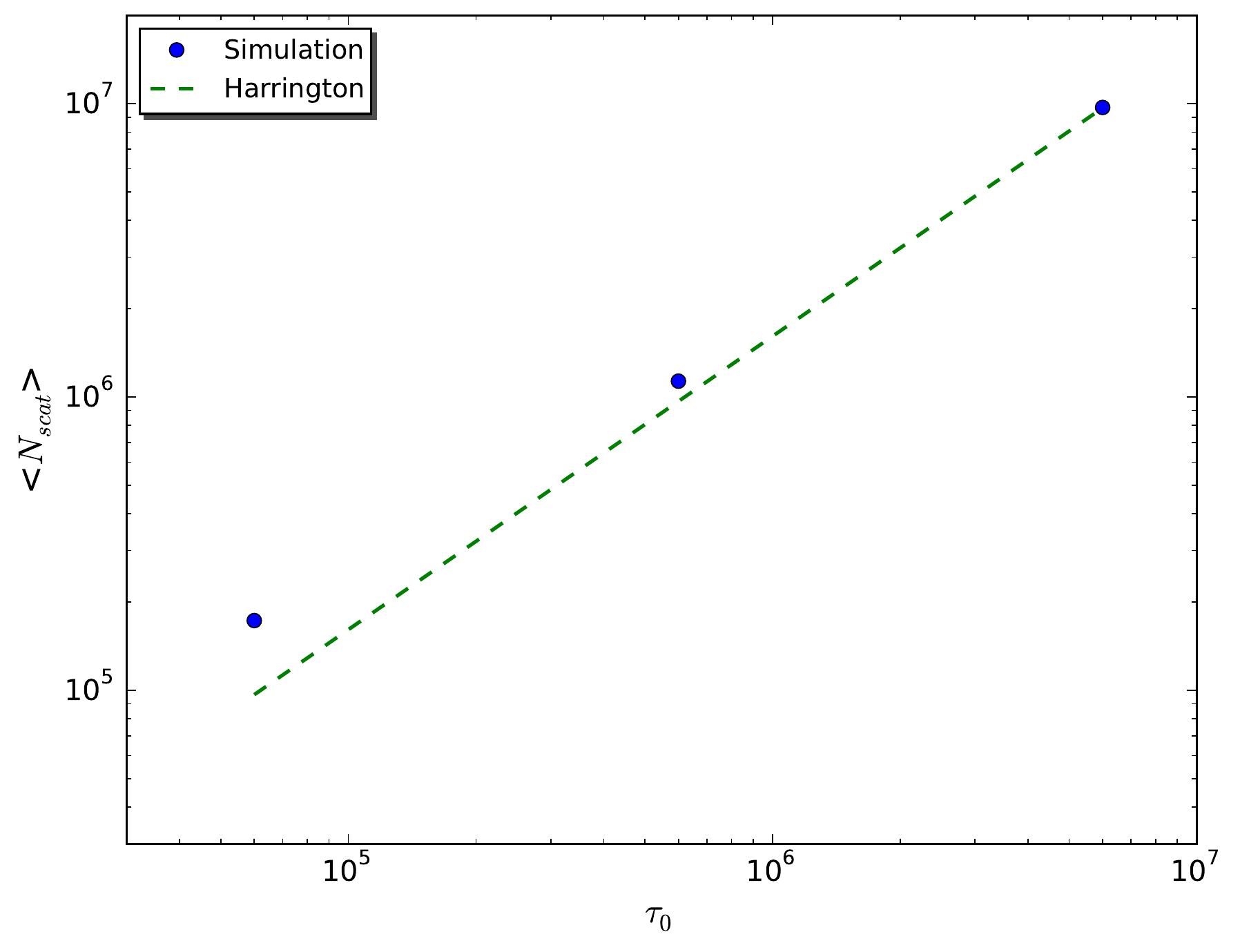}\caption{\label{fig:Number-of-scattering} 
Average number of scatterings until the photons escape the system. The results from our Ly$\alpha$ scattering code are shown for $\tau_0=6\times 10^4$, $6\times 10^5$, and $6\times 10^6$ as the blue dots. The analytic solution from Equation~(\ref{eq:num_scat}) is shown as the green dashed line.}
\end{figure}

The scattering count also has an analytic solution. \citet{1973MNRAS.162...43H} derived that the mean scattering count is
\begin{equation} \label{eq:num_scat}
\left< N_{\rm scat} \right>=1.612\times\tau_{0}.
\end{equation}
In Figure \ref{fig:Number-of-scattering}, we compare the scattering count from our Ly$\alpha$ scattering code to the analytic solution. Our result converges to the analytic solution from Equation~(\ref{eq:num_scat}) in the high-$\tau_0$ limit. 

\subsection{Static Homogeneous Sphere}

\begin{figure}
\includegraphics[width=0.4\paperwidth]{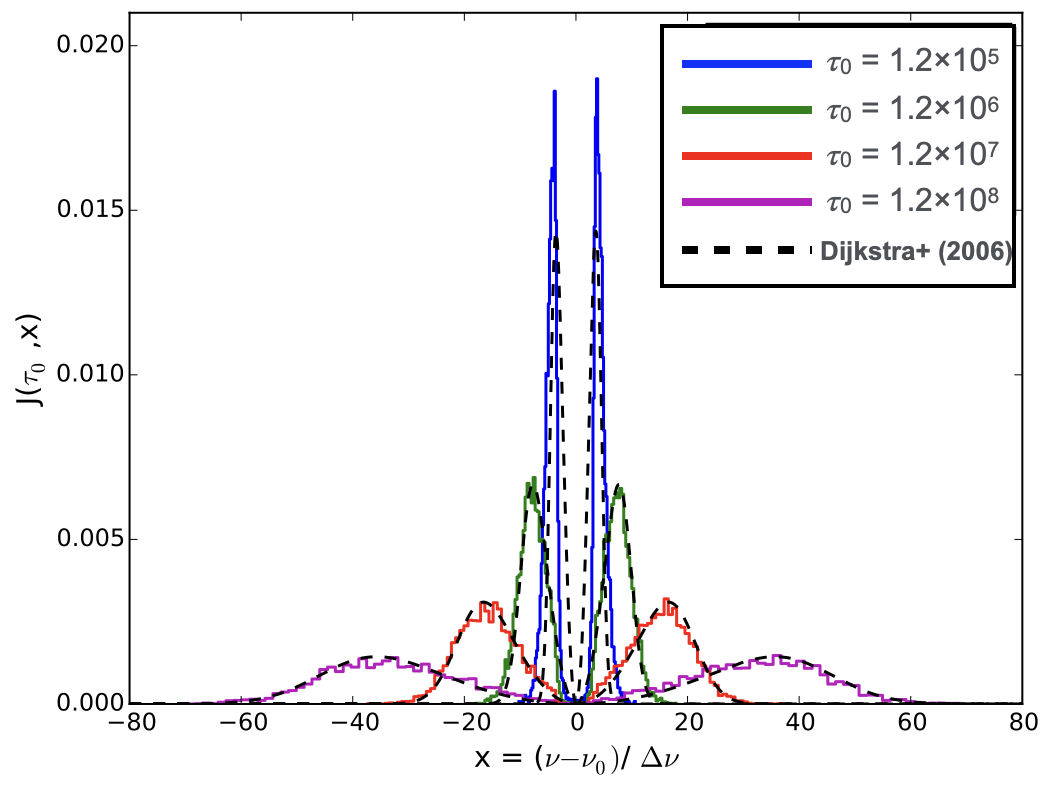}
\caption{\label{fig:Static_sphere} Emergent SED from the homogeneous and isothermal static sphere for $\tau_0=1.2\times10^5$, $1.2\times10^6$, $1.2\times10^7$, and $1.2\times10^8$, shown as the solid lines. The analytic solution from \citet{2006ApJ...649...14D} is shown as the black dashed lines.}
\end{figure}

\begin{figure*}
\includegraphics[width=0.4\paperwidth]{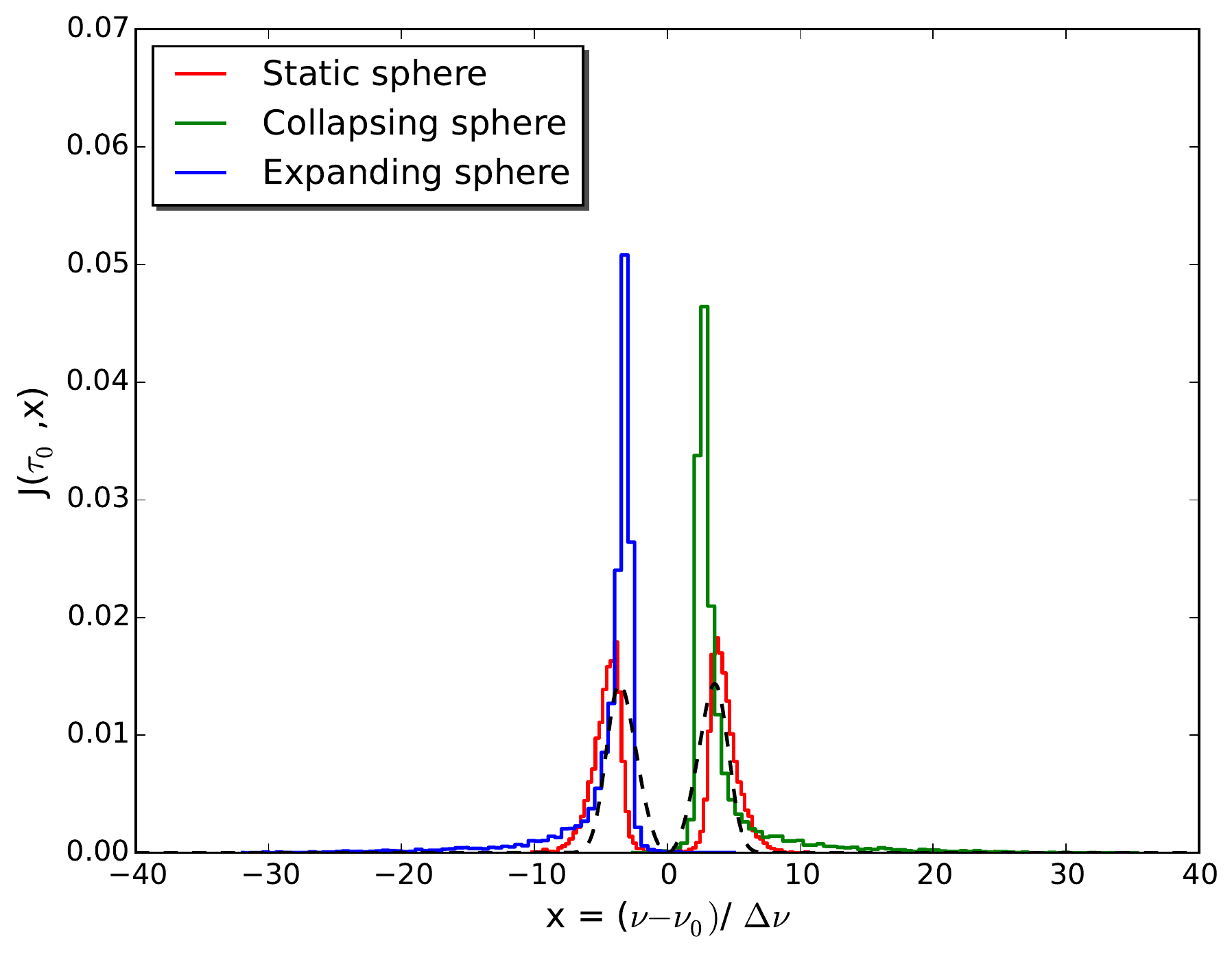}
\includegraphics[width=0.4\paperwidth]{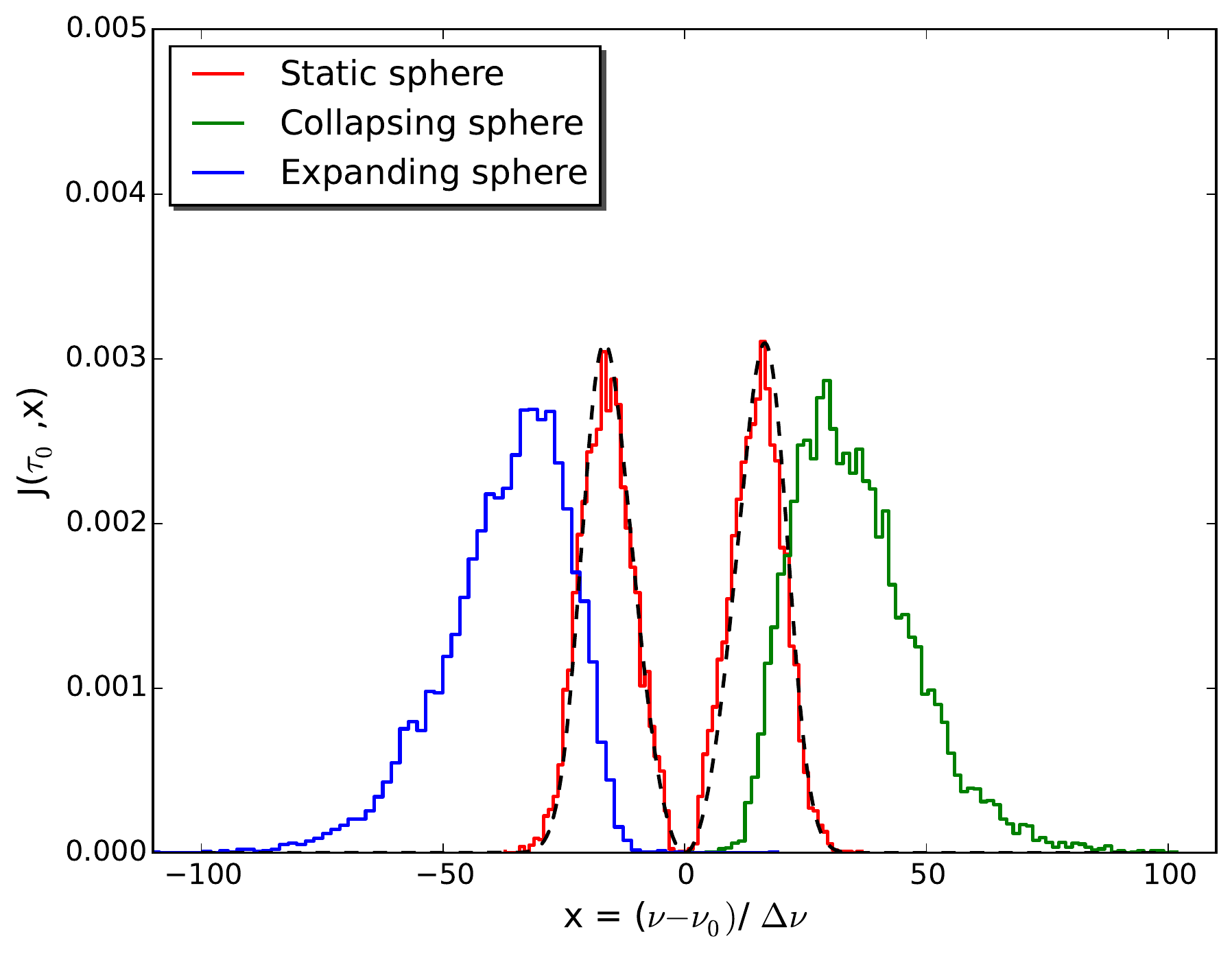}
\caption{\label{fig:expnad-te5} The emergent spectrum of the Ly$\alpha$ source located at the center of the static, expanding, and collapsing homogeneous and isothermal spheres. The gas temperature is set to $T = 10^4$ K, and the HI density is set so that the optical depth from the center to the surface is  $\tau_{0}=1.2\times10^5$ and $1.2\times10^{7}$ in the left and right panels, respectively. For each density, we explore the cases of collapsing (green solid), static (red solid), and expanding (blue solid) sphere defined by the maximum radial speed at the surface of $V_{\rm max}=-200$, $0$, and $200$ km ${\rm s}^{-1}$, respectively. The analytic solution of \citet{2006ApJ...649...14D} for the static case is shown as the black dashed line.}
\end{figure*}

We test the code for a static homogeneous sphere. We generate a cubic volume of $256^{3}$ cells of homogeneous density, peculiar velocity, the ionization fraction, and the temperature fields. The HI column density is varied as $N_{\rm HI} =2\times10^{18},$ $2\times10^{19},$ $2\times10^{20}$, and $2\times10^{21}\;{\rm cm}^{-2}$ which correspond to the line center optical depths of $\tau_{0}=1.2\times10^{5},$ $1.2\times10^{6},$ $1.2\times10^{7},$ and $1.2\times10^{8}$, respectively; the line center optical depth is defined as the optical depth between the center of the sphere and the surface. The density field is generated to meet the intended column density. The sphere size is set to 10 kpc. The small sphere size allows the code to ignore the effect of cosmic expansion. We place a monochromatic Ly$\alpha$ source at the center of the sphere and set the temperature to $T=10^{4}$ K everywhere.

\citet{2006ApJ...649...14D} derived the analytic solution for this configuration. 
In the solution, angle-averaged mean intensity is given by 
\begin{equation}
J(\tau_0, x) =\frac{\sqrt{\pi}}{\sqrt{24}a\tau_{0}}\frac{x^{2}}{1+\cosh[\sqrt{{2\pi^{3}}/{27}}(\vert x\vert^{3}/a\tau_{0})]}.
\end{equation}

The test results are shown in Figure \ref{fig:Static_sphere}. Except the case of $\tau_{0}=1.2\times10^{5}$, the spectral energy distribution (SED) from the code accurately reproduces the analytic solution with only small deviations from statistical fluctuations. Because the analytic solution is derived for the optically thick limit, $\tau_{0}=1.2\times10^{5}$ shows a greater deviation than other cases with larger optical depths, as was also reported many times by previous studies \citep{2006ApJ...649...14D,2006A&A...460..397V,2007A&A...474..365S,2009ApJ...696..853L}.

\subsection{Homogeneous sphere with Hubble-like flow}

On top of the case of a static homogeneous introduced above, we apply a radially outward motion defined by $\mathbf{v}_{H}=V_{\rm max}\mathbf{r}/R_{\rm max}$, where $R_{\rm max}=10\;{\rm kpc}$ is the radius of the sphere. We set the column density from the center to the surface as $N_{\rm HI}=2\times10^{18}$ and $2\times10^{20}\,{\rm cm}^{-2}$ corresponding to $\tau_0 = 1.2\times10^5$ and $1.2\times10^7$, respectively.  For each density, we set three different maximum velocities at the surface, $V_{\rm max}=-200$, $0$, and $200~{\rm km}~{\rm s}^{-1}$, to explore cases of a collapsing, static, and expanding sphere. The results for this configuration are available from previous works \citep{1999ApJ...524..527L,2002ApJ...578...33Z,2006ApJ...649...14D,2006ApJ...645..792T,2006A&A...460..397V,2007A&A...474..365S,2009ApJ...696..853L}.

\begin{figure}
\includegraphics[width=0.4\paperwidth]{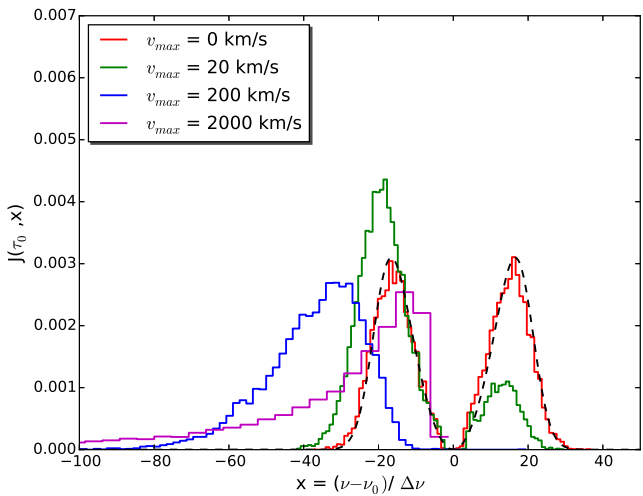}\caption{\label{fig:expand-te7-vvary} The emergent spectrum of the Ly$\alpha$
scattering in a static, expanding homogeneous and isothermal sphere. $N_{\rm HI}=2\times10^{20}\;{\rm cm}^{-2}$, $\tau_{0}=1.2\times10^{7}$, and
$T =10^4$ K.}
\end{figure}

The results are shown in Figure \ref{fig:expnad-te5}. In the expanding sphere, the red side of the spectrum is enhanced while the blue part is completely suppressed. This is because all the photons that are emitted on the blue side are scattered in the outskirts, where the expansion is fast enough to shift the frequency to the resonance. The collapsing sphere works oppositely and enhances the blue part of the spectrum. In Figure~\ref{fig:expand-te7-vvary}, we fix the HI column density to $N_{\rm HI}= 2\times10^{20}~{\rm cm}^{-2}$ and vary the expansion velocity ($V_{\rm max}= 0, 20,~200,$ and $~2000~{\rm km}~{\rm s}^{-1}$). As we increase $V_{\rm max}$ from $0$ to $2000~{\rm km}~{\rm s}^{-1}$, the blue part is suppressed, and the red peak is extended to longer wavelengths. The red peak shifts toward longer wavelengths as $V_{\rm max}$ increases up to 200 km/s, but it shifts back toward the line center above a certain threshold value, as can be seen from the 2000 km/s case because the steep velocity gradient allows more photons to escape before redshifting further \citep{2009ApJ...696..853L}. Similar results can be found in Figure 8 of \cite{2009ApJ...696..853L} and in the right panels of Figures 2 and 3 in \citet{2002ApJ...578...33Z}.

\begin{figure*}
\begin{center}
\includegraphics[width=0.75\paperwidth]{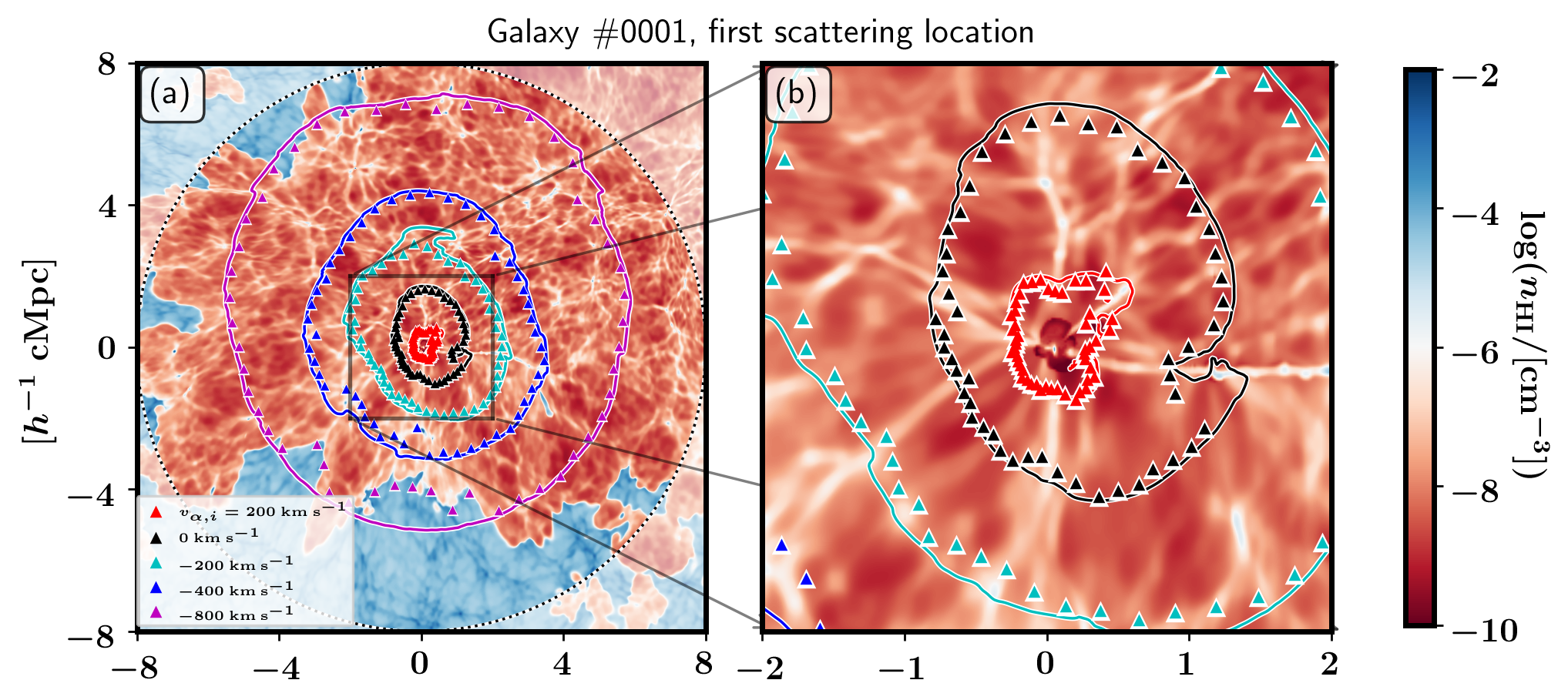}
\includegraphics[width=0.75\paperwidth]{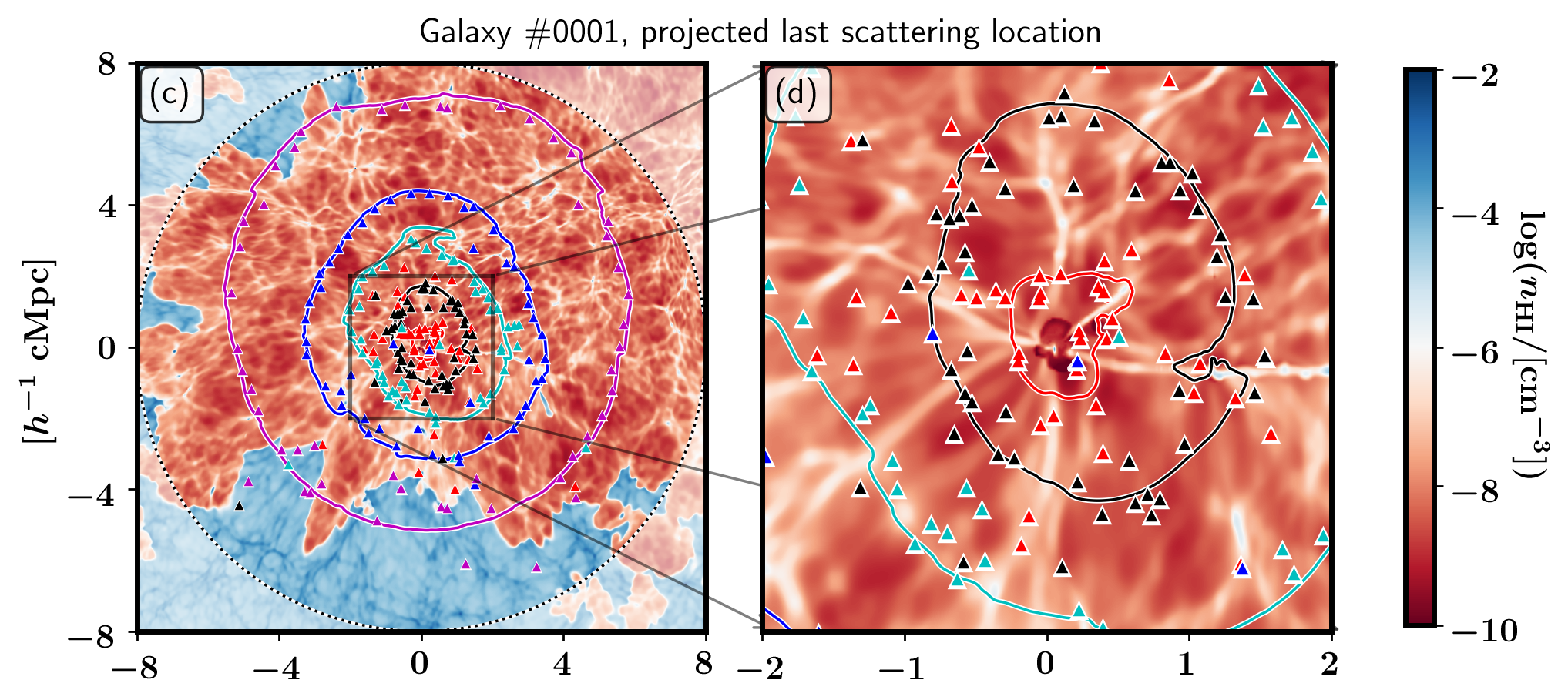}
\caption{\label{fig:plane} HI density map of the $xy$ plane containing galaxy \#0001 and the first (last) scattering location of the Ly$\alpha$ photons that are initially emitted on that plane in the upper (lower) panels. We generate a photon for every 1 degree from the initial azimuthal angle of $\phi=0^{\circ}$ to $359^\circ$ with zero latitude angle from the plane. The $x$- and $y$-axis ticks are all in units of $h^{-1}~{\rm Mpc}$. The left panels shows a slice of the entire $16h^{-1}~{\rm cMpc}$ box used for the Ly$\alpha$ RT calculation, and the right panels shows a zoomed-in central region that is $4$ $h^{-1}~{\rm cMpc}$ on a side. The black dotted lines in the left panels mark the boundary of the RT calculation ($8$ $h^{-1}~{\rm Mpc}$ from the source), where we assume the photon has escaped the system and sample the photon information. The red/blue regions on the map generally corresponds to ionized/neutral parts of the IGM. The red, black, cyan, blue, and magenta triangle symbols show the scattering location for the photons with initial wavelengths $v_{\alpha,i}=200$, $0$, $-200$, $-400$, and $-800~{\rm km}~{\rm s}^{-1}$, respectively. The line contours connect $r_s$ from Equation~(\ref{eq:scatter}) for each direction from the galaxy, where we expected the photons to be scattered for the first time. The last scattering positions are the projected locations on the $xy$ plane. }
\end{center}
\end{figure*}

\section{Application to Reionization Simulation Data} \label{sec:CoDaII}

\begin{figure*}
\begin{center}
\includegraphics[width=0.75\paperwidth]{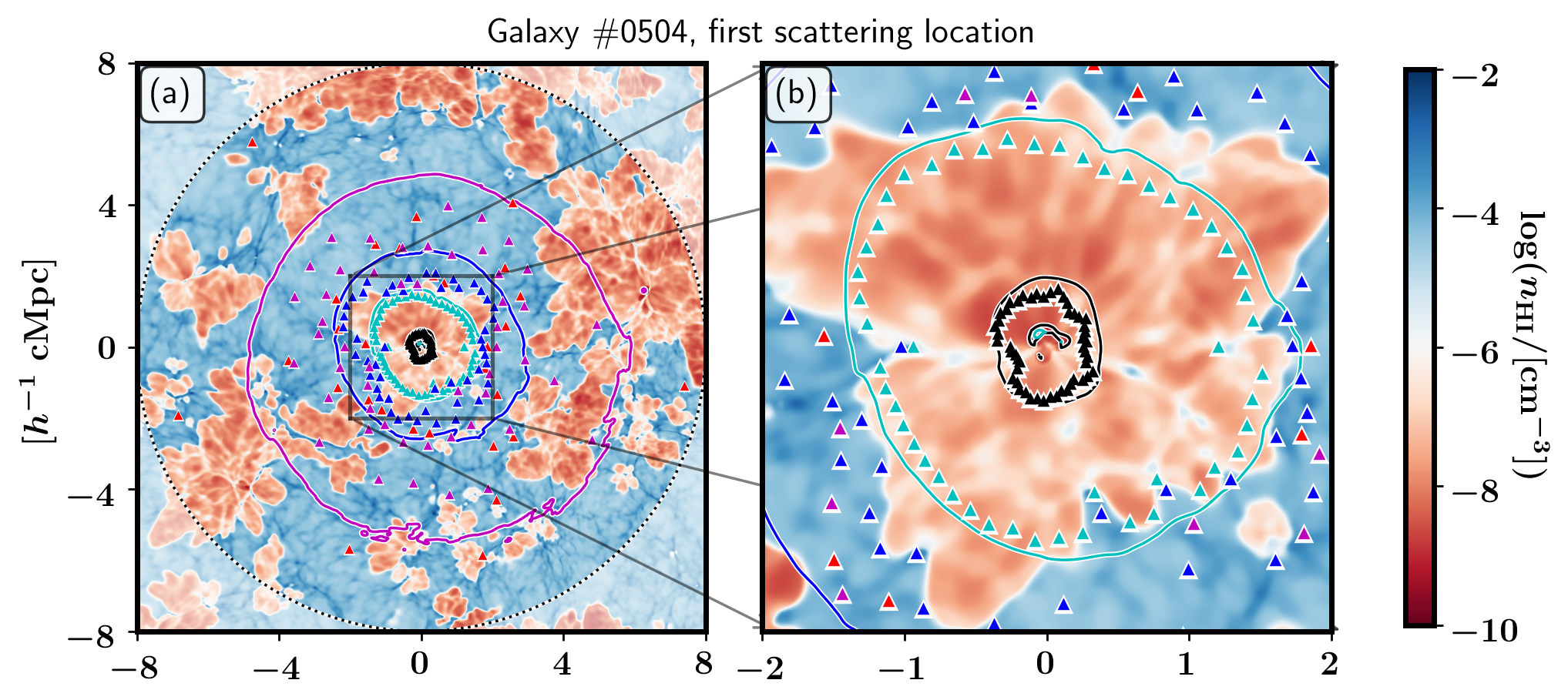}
\includegraphics[width=0.75\paperwidth]{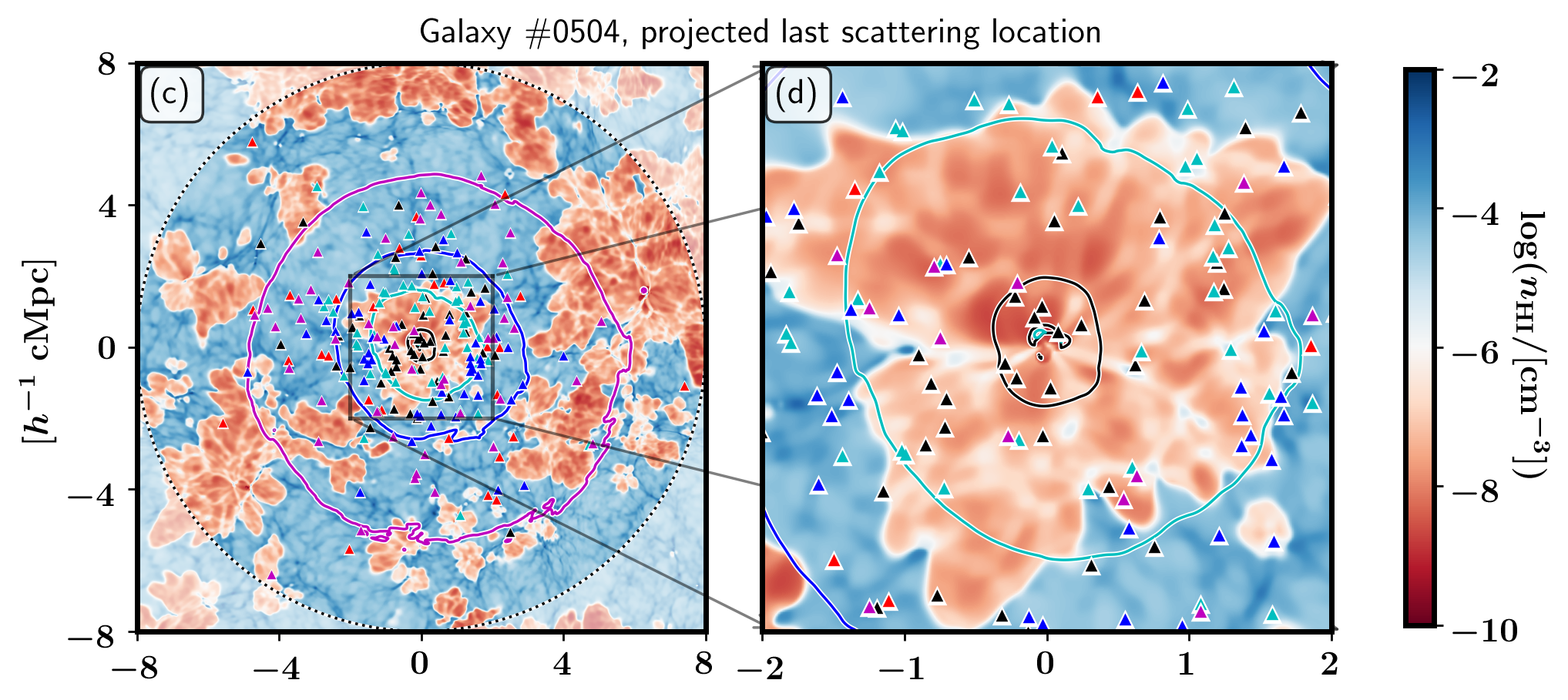}
\caption{\label{fig:plane2} Same as Figure~\ref{fig:plane} but for galaxy \#0504.}
\end{center}
\end{figure*}

We apply our Ly$\alpha$ RT code to the CoDaII simulation dataset. As introduced above, CoDaII is dedicated to reproducing the IGM during the epoch of reionization by simulating the formation of early galaxies and the ionizing radiation from them. The output data include the ionization fraction ($\chi$), density ($\rho$), peculiar velocity ($\bold{v}_{\rm pe}$), and temperature ($T$) fields of gas on a $4096^3$ mesh of a $64h^{-1}~{\rm cMpc}$ box and basic galaxy properties including absolute UV magnitude ($M_{\rm UV}$) and halo mass ($M_h$). Our goal is to find how the observed Ly$\alpha$ line shape would change if we collected the scattered Ly$\alpha$ emission around a UV-bright galaxy along with the unscattered light coming directly from the galaxy in high-$z$ integral-field-unit (IFU) surveys. 

We limit the scope of this work to the SED of the scattered light from two galaxies in the $z=7$ snapshot, where the IGM is $50\%$ ionized in the simulation. The two galaxies that we analyze are the first and 504th brightest galaxies in the snapshot, which we name as galaxy \#0001 and galaxy \#0504. The UV magnitudes of these galaxies are $M_{\rm UV}=-23.1$ and $-19$, and their total masses are $M_h=1.1\times 10^{12}$ and $6.7\times 10^{10}~M_\odot$, respectively. The former is surrounded by a relatively large HII region ($\gtrsim 5$ cMpc), while the latter is surrounded by a smaller one ($\sim 2$ cMpc). The latter case is meant to represent the early stage of reionization, while the former represents the late stage or the post-reionization regime. At $z=7$ in the simulation, the snapshot has a mixture of both cases, making it suitable for exploring both regimes from a single snapshot. We shall provide a more comprehensive analysis with more galaxies for other physical quantities such as the surface brightness profile in our future work.

For each sample galaxy, we trim out the $16h^{-1}~{\rm cMpc}$ box with the galaxy at the center. We calculate the gas-density-weighted mean peculiar velocity within $r_{200}$ and subtract it from the entire velocity field to work in the source galaxy frame. Then, we generate photons at $r_{200}$ from the source galaxy and initialize the photons to propagate radially outward\footnote{We shall test the case of radial emission against one of nonradial emission in Section~\ref{sec:red_double_peak_model} and in the appendix.}. Given the limited spatial resolution of the simulation below the circumgalactic scales, we do not attempt to simulate RT within $r_{200}$ and instead focus on the scattering process in the IGM. When a sample photon reaches $8h^{-1}~{\rm cMpc}$ from the source, we assume it has escaped the system, and we sample the final frequency ($\nu_{\rm obs}$) and the transverse distance to the source $r_\perp$, as described in Section~\ref{sec:cosmo_effect}. 

Since the gas temperature is not constant in the simulation, the dimensionless frequency $x$ is not convenient for describing the results. Thus, we instead use the wavelength offset from Ly$\alpha$ in the velocity unit defined as 
\begin{equation}
v_{\alpha}\equiv -c\frac{\nu_{\rm obs}-\nu_0}{\nu_0}.
\end{equation}
In this unit, $1$ {\AA} roughly corresponds to $250~{\rm km}~{\rm s}^{-1}$ at $z=7$.

\subsection{Scattering Location} \label{sec:SL}

As demonstrated in \citet[][hereafter P21]{2021ApJ...922..263P}, the residual HI density in the HII regions during the epoch of reionization is generally high enough to keep the IGM opaque at the Ly$\alpha$ resonance ($n_{\rm HI} \gtrsim 10^{-9}~{\rm cm}^{-3}$) even in the near-zone of UV-bright galaxies \citep[see also, e.g.,][]{2008MNRAS.391...63I}. Thus, a photon emitted on the blue side of the resonance in the IGM frame will eventually redshift to the resonance after the propagation distance $r_s$ given by 
\begin{equation}\label{eq:scatter}
Hr_s+v_{\alpha,i}+v_{{\rm pe},r}=0,
\end{equation}
where $H$ is the cosmic expansion rate, $v_{\alpha,i}$ is the initial wavelength of the photon at emission, and $v_{{\rm pe},r}\equiv\hat{\bold{r}}\cdot \bold{v}_{\rm pe}$ is the radial peculiar motion of the IGM. Unless there is a highly neutral region on the way, the photon would propagate freely until reaching $r_s$ from the source and be scattered for the first time after its emission. 

Due to the gravitational field of the source galaxy, the IGM generally infalls toward the galaxy (i.e., $v_{{\rm pe},r}<0$). As a result, some photons that are emitted on the red side of the resonance in the source frame can be on the blue side in the IGM frame if $v_{\alpha,i}<-v_{{\rm pe},r}$. These photons would also redshift to the resonance after propagating a distance $r_s$ from the source. 

According to P21, $v_{{\rm pe},r}(r)$ is given approximately by $-(GM_h/r)^{0.5}$ with some variations due to the gravitational field of the neighboring density structures. Thus, $r_s$ for a given $v_{\alpha,i}$ forms a near-spherical ``first scattering" surface surrounding the source galaxy.

\subsubsection{Large HII bubble case} \label{sec:LBC}

In a large HII region, most photons starting blueward of the resonance in the IGM frame ($v_\alpha+v_{{\rm pe},r}<0~{\rm km}~{\rm s}^{-1}$) would propagate uninterrupted until they redshift to the resonance. In this case, we expect Equation~(\ref{eq:scatter}) to accurately give the first scattering location. We test this hypothesis for galaxy \#0001, which is surrounded by a relatively large HII region extending beyond $5h^{-1}~{\rm cMpc}$ from the galaxy in most directions, as shown in Figure~\ref{fig:plane}.

We show the first scattering surface on the $xy$ plane for $v_{\alpha,i}$ $=-800$, $-400$, $-200$, $0$, and $200~{\rm km}~{\rm s}^{-1}$ as line contours in Figure~\ref{fig:plane}. The contour is highly circular because the IGM infall motion is nearly isotropic, and it is larger for smaller $v_{\alpha,i}$ because bluer photons travel a greater distance $r_s$ to reach the resonance. The initially red photons with $v_{\alpha,i}=200~{\rm km}~{\rm s}^{-1}$ also form the first scattering contour because the gravitational infall motion around galaxy \#0001 exceeds $200~{\rm km}~{\rm s}^{-1}$, making them blue-side photons in the IGM frame\footnote{See Section 3.1 of P21 for the detailed analysis of the infall motion.}.

We run our Ly$\alpha$ RT code for the photons that are initially emitted in the $xy$ plane of the source and show their first scattering locations as triangles in Figure~\ref{fig:plane}. The figure shows the scattering locations for 360 photons with their initial azimuthal angle between $0^\circ$ and $359^\circ$ and zero latitude angle from the plane. The first scattering location generally coincides with the $r_s$-contour except for some downward directions (i.e., near the $-y$ direction) for the photons with $v_{\alpha,i}=-800~{\rm km}~{\rm s}^{-1}$, which enter the neutral region and are scattered before reaching the contour. Other than this case, all the scattering locations fall slight inside the contour with a small offset of $\lesssim 0.1h^{-1}~{\rm cMpc}$. This offset occurs because photons are scattered when the optical depth exceeds one, while the $r_s$-contour marks the peak of the IGM opacity, which happens slightly later. The offset is generally small compared to the value of $r_s$, indicating it is a good description for the first scattering location in HII regions. The first scattering location, in principle, has a distribution according to the optical depth distribution of the propagation. However, almost all the photons are scattered within a thin surface near the $r_s$-contour because the optical depth remains small until the photon approaches the contour, where the optical depth rises steeply. This behavior of the optical depth is described in detail in Figure 4 of P21.

The last scattering locations shown in the lower panels of Figure~\ref{fig:plane} also coincide with the $r_s$-contours, indicating that the photons do not travel far between the first and the last scatters. The agreement for the redward emission ($v_{\alpha,i}=200~{\rm km}~{\rm s}^{-1}$) is not as good as that for the bluer emissions, but it is still much better than in the small HII bubble case (Fig.~\ref{fig:plane2}), as we will describe below. We thus conclude $r_s$ well describes the last scattering location of escaped photons, as well as the first scattering location, in large HII regions.

\subsubsection{Small HII bubble case}

Galaxy \#0504 has an HII region of roughly $1h^{-1}~{\rm cMpc}$ (see Figure~\ref{fig:plane2}), which is smaller than that surrounding galaxy \#0001. In such a small HII region, most Ly$\alpha$ photons would enter the HI region before redshifting to or away from the resonance. Thus, the damping-wing cross section of the HI gas creates a large opacity for these photons, and scatters most of them before they reach $r_s$.

In Figure~\ref{fig:plane2}, the first scattering location agrees well with the $r_s$-contour for $v_{\alpha,i}=0$ and $-200~{\rm km}~{\rm s}^{-1}$, of which the $r_s$-contour lies within the HII region. However, the photons with $v_{\alpha,i}=-400$ and $-800~{\rm km}~{\rm s}^{-1}$ are scattered well inside the $r_s$-contour because the HI gas scatters the photons before they redshift to the resonance with its damping-wing opacity. The first scattering location of the $v_{\alpha,i}=-400~{\rm km}~{\rm s}^{-1}$ case closely follows the boundary of the HII region, indicating that the optical depth is rising steeply there for these photons. On the other hand, the first scattering location of the $v_{\alpha,i}=-800~{\rm km}~{\rm s}^{-1}$ case is more scattered between the HII region boundary and its $r_s$-contour. This is because these photons enter the HI region when their frequencies are relatively far from the line center, where the Ly$\alpha$ cross section is low, and the IGM opacity rises more gradually, resulting in the scattering probability being distributed more extensively in space.

The photons emitted at $v_{\alpha,i}=200~{\rm km}~{\rm s}^{-1}$ do not encounter the resonance because the peculiar infall velocity of this galaxy is $\sim 150~{\rm km}~{\rm s}^{-1}$, which is not strong enough to put those photons blueward of resonance in the IGM comoving frame. 43\% of these photons with $v_{\alpha,i}=200~{\rm km}~{\rm s}^{-1}$ escape the RT simulation volume unscattered, while the rest are scattered within the HI region due to the damping-wing opacity.

The first and last scattering locations are much less correlated than in the case of a large HII region because a significant fraction of the photons are additionally scattered in the surrounding HI region. Therefore, the $r_s$-contour is a poor description for the scattered light in a small HII region like this one.

\begin{figure*}
\begin{center}
\includegraphics[width=0.4\paperwidth]{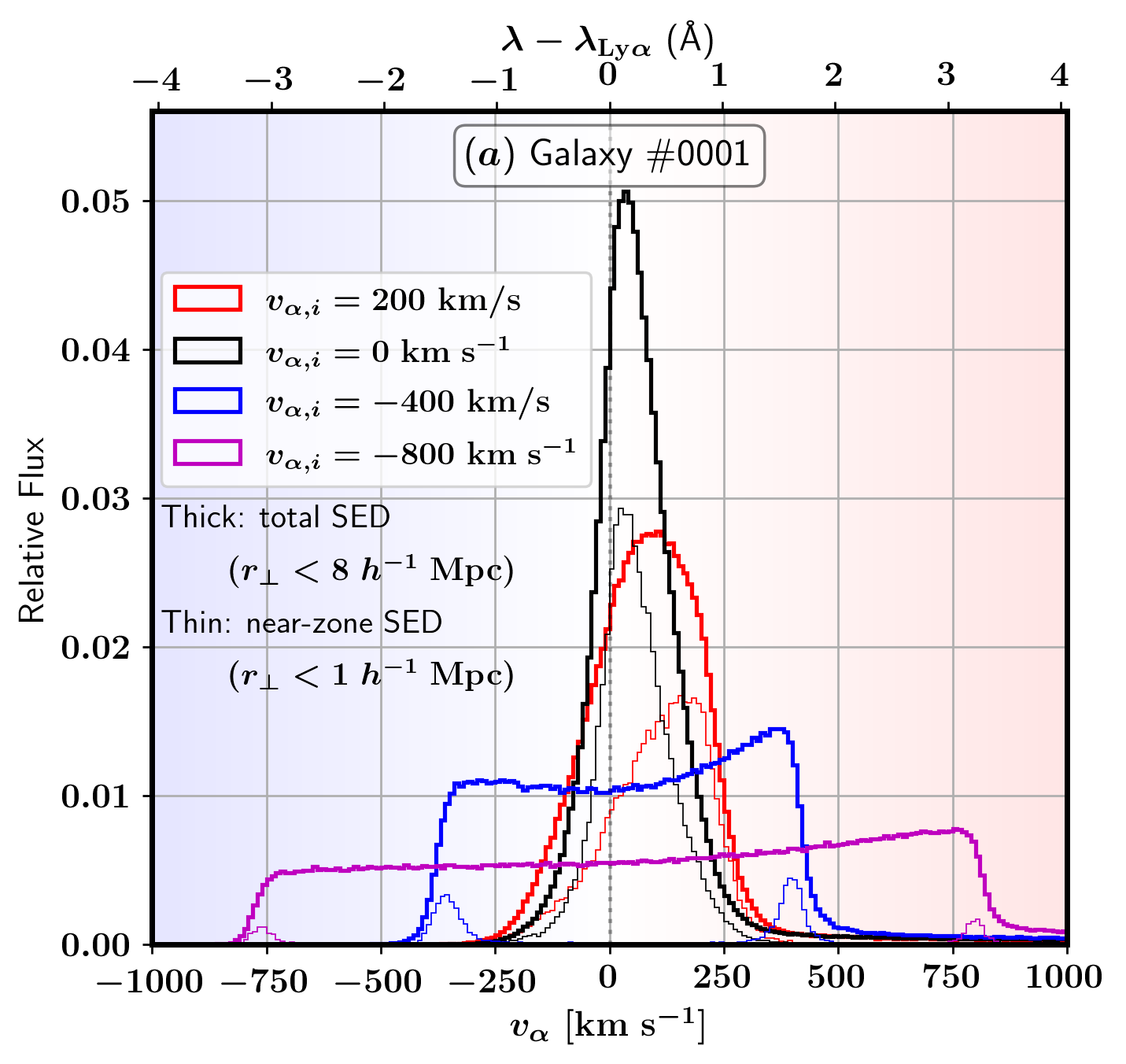}
\includegraphics[width=0.4\paperwidth]{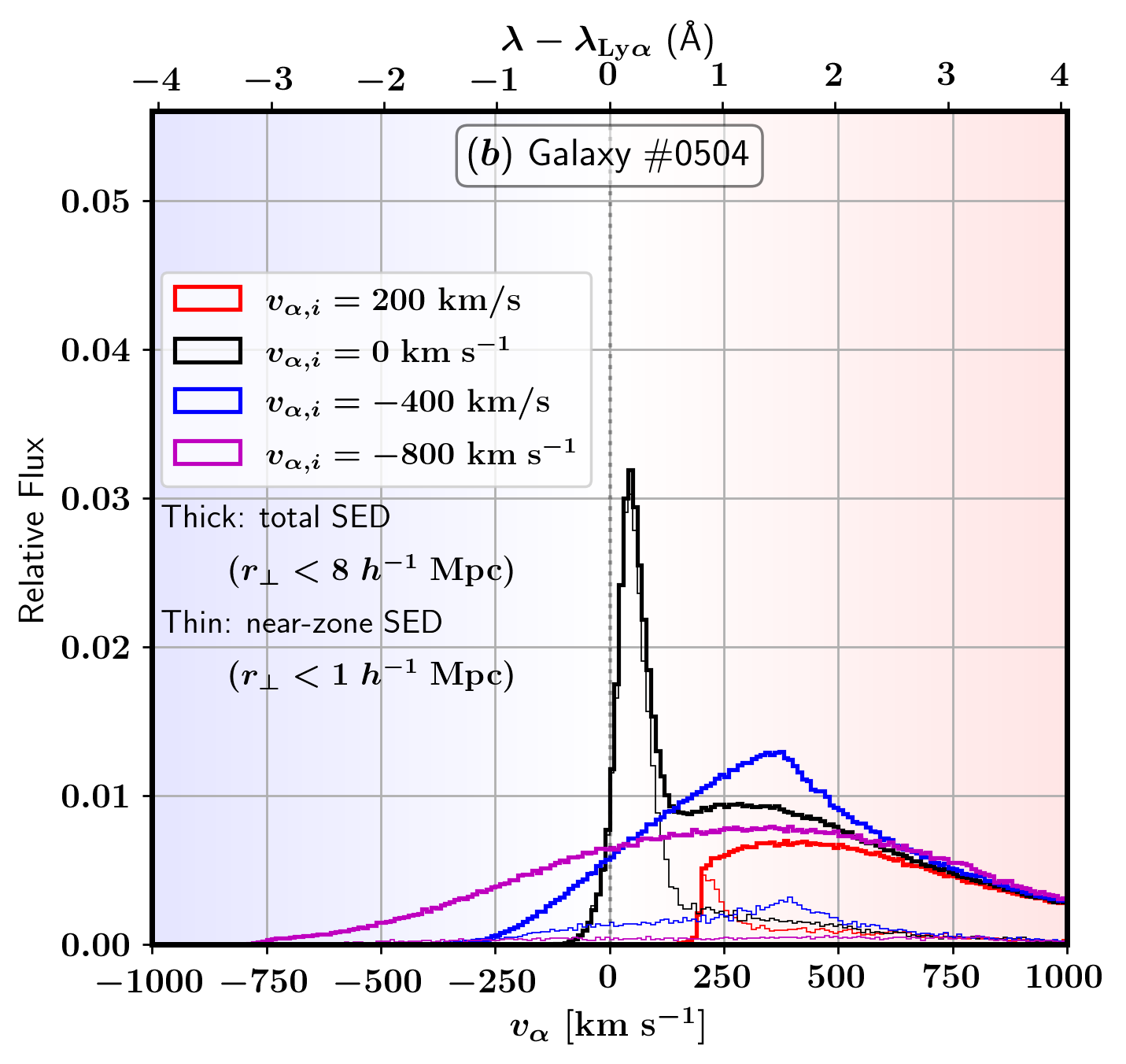}
\caption{\label{fig:spec} Emergent SED of the scattered photons with $v_{\alpha,i}=-800,$ $-400,$ $-200,$ $0$ and $200~{\rm km}~{\rm s}^{-1}$ shown as magenta, blue, cyan, black, and red histograms, respectively, in the source frame. The results for galaxies \#0001 and \#0504 are shown in the left and right panels, respectively. The thick lines are for the flux from all the sampled photons sampled at $r_\perp<8h^{-1}~{\rm cMpc}$, and the thin lines are for those sampled at $r_\perp<1h^{-1}~{\rm cMpc}$. For galaxies \#0001 and \#0504, 9\% and 43\% of the photons with $v_{\alpha,i}=200~{\rm km}~{\rm s}^{-1}$ escape the system unscattered and are excluded from the histogram, respectively.}
\end{center}
\end{figure*}

\subsection{Scattered Light SED: Monochromatic Sources} \label{sec:spectrum}

Before looking into the scattered light for extended emission profiles, we first explore monochromatic cases, where the photons are initially emitted at a fixed wavelength of $v_{\alpha,i}=-800,~-400,~-200,~0,$ or $200~{\rm km}~{\rm s}^{-1}$. These monochromatic cases are not realistic, but they are useful for understanding the relation between input and output spectra, because the output spectrum from an arbitrary input spectrum can be constructed from superpositions of the monochromatic cases.

We initialize the photons at random locations on the sphere of $r_{200}$ with radially outward propagation direction, assuming the source emissivity is isotropic. We obtain the SED by taking the probability distribution of $v_{\alpha}$ of all the sampled photons, effectively averaging the observations from all possible viewing angles. We show these results in Figure~\ref{fig:spec}. We show the SED for all the sampled photons at $r_\perp \le 8h^{-1}~{\rm cMpc}$ (or $4.5~{\rm arcmin}$) and for a fraction of photons that are sampled within $r_\perp = 1h^{-1}~{\rm cMpc}$ (or $0.6~{\rm arcmin}$) from the source galaxy.  This way, we account for the impact of having a finite light-collecting area on the observed SED. In practical observations, it would be difficult to collect the scattered light beyond $\sim 1~{\rm arcmin}$ because of the radiation from other galaxies in the field and the sky noise \citep{2011ApJ...739...62Z}. We shall refer to the former case as the ``total" SED and the latter as the ``near-zone" SED. As in the previous sections, galaxies \#0001 and \#0504 represent large and small HII bubbles around the source galaxies.

\subsubsection{Large HII bubble case}
The total SED of galaxy \#0001 (thick histogram in the left panel of Figure~\ref{fig:spec}) is narrowly peaked at the resonance for $v_{\alpha,i}=0$ and $200~{\rm km}~{\rm s}^{-1}$ and becomes wider for smaller $v_{\alpha,i}$'s (i.e., shorter initial wavelengths). For $v_{\alpha,i}=-400$ and $-800~{\rm km}~{\rm s}^{-1}$, the emergent SED shape is similar to a top hat extending from $v_\alpha=v_{\alpha,i}$ to $-v_{\alpha,i}$. 

The near-zone SED is similar to the total SED in shape with mildly lower ($\sim 30\%$) intensity for $v_{\alpha,i}=0$ and $200~{\rm km}~{\rm s}^{-1}$. For the bluer emission cases ($v_{\alpha,i}=-800$ and $-400~{\rm km}~{\rm s}^{-1}$), nearly all the emission around the line-center from the total SED is lost, and only the red and blue tips of the distribution are captured in the near-zone SED.

\subsubsection{Small HII bubble case}
In the case of a small HII bubble (right panel of Figure~\ref{fig:spec}), the emergent SED is more extended redward than in the case of a large HII bubble. For $v_{\alpha,i}=-800$ and $-400~{\rm km}~{\rm s}^{-1}$, the blue end of the total SED coincides with $v_{\alpha,i}$, as it was in the case of the cae of a large HII bubble, but the SED is much weaker on the blue side. Instead, the SED is substantially more extended redward going beyond $1000~{\rm km}~{\rm s}^{-1}$. 

The near-zone SED is also much weaker and extended redward.  For $v_\alpha=0$ and $ 200~{\rm km}~{\rm s}^{-1}$, the blue end of the near-zone SED lies on the total SED, but the redward emission is mostly not included in the near-zone. The near-zone SED is weaker at all wavelengths, and the double-peaked feature seen in the case of a large HII bubble does not appear in this case. 

\begin{figure*}
\begin{center}
\includegraphics[width=0.7\paperwidth]{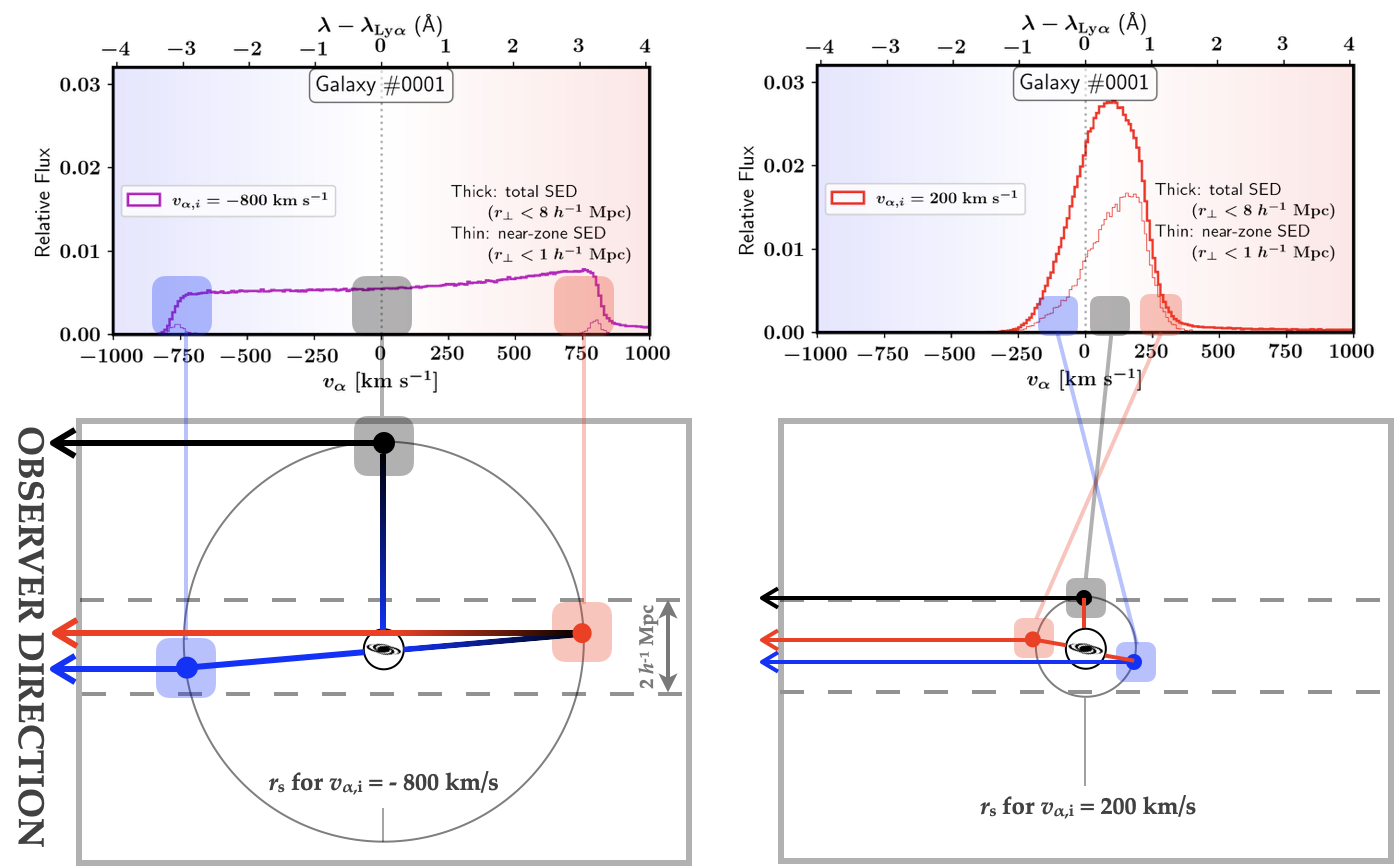}
\caption{\label{fig:LSP} Schematic of the photon paths for $v_{\alpha,i}=-800$ (left) and $200~{\rm km}~{\rm s}^{-1}$ (right) emitted from galaxy \#0001 (lower panels) and the corresponding final wavelengths seen by the observer (upper panels). The scattering location is given by Eq.~\ref{eq:scatter} as elaborated in Sec.~\ref{sec:LBC}. The color gradation in the arrows in the left panel describes the cosmological redshift due to the extra propagation distance from scattering. The color discontinuity in the arrows in the right panels describes the change in frequency during the scattering due to the peculiar velocity of the scattering atom. The pair of dashed lines describe the near-zone defined as the region within $r_\perp=1h^{-1}~{\rm Mpc}$ (or 0.6 arcmin) from the source galaxy. }
\end{center}
\end{figure*}

\subsubsection{Physical Explanation} \label{sec:explanation}

Here, we provide a physical explanation for the scattered light SEDs of the monochromatic cases shown above. We first describe the SEDs for the case of a large HII region (left panel of Figure~\ref{fig:spec}). Then, the case of a small HII region (right panel of Figure~\ref{fig:spec}) can be understood by considering additional scatterings in the surrounding HI region. 

The schematic in the left panel of Figure~\ref{fig:LSP} describes the scattering location and the paths of photons with $v_{\alpha,i}= -800~{\rm km}~{\rm s}^{-1}$, which are emitted on the far blue side of Ly$\alpha$. These photons travel a relatively large distance until being scattered toward the observer. The peculiar infall velocity of the IGM is roughly $(GM_h/r_s)^{0.5}\sim 70~{\rm km}~{\rm s}^{-1}$ at the scattering surface, which is small relative to the initial offset from Ly$\alpha$ ($800~{\rm km}~{\rm s}^{-1}$). Thus, the IGM infall motion is less important for shaping the SED than the cosmological redshift during propagation. We use a color gradation in the arrows to illustrate how the cosmological redshift effect makes photons emitted at the same wavelength end up at different wavelengths. If a photon is initially emitted toward the observer, its path length would be similar to the direct distance to the observer, and the observed wavelength would be similar to the initial wavelength (i.e., $v_{\alpha}\approx v_{\alpha,i}$). When the photon is emitted in the opposite direction and later scattered toward the observer at the scattering surface as in the red path, the path length increases by $2r_s$, redshifting the photon to $v_{\alpha}\approx -v_{\alpha,i}=800~{\rm km}~{\rm s}^{-1}$. Likewise, the black path shows that the photons initially emitted perpendicular to the observer's direction would travel an extra distance of $r_s$ and be observed near the line center.

These example paths show that the emergent wavelength of the scattered photon is given by $v_{\alpha}=v_{\alpha,i} \cos{\theta}$, where $\theta$ is the angle between the initial and final photon directions. This explains why the total SEDs for $v_{\alpha,i}= -400$ and $-800~{\rm km}~{\rm s}^{-1}$ have the shape of a top-hat extending from $v_{\alpha}=v_{\alpha,i}$ to $-v_{\alpha,i}$: $\cos{\theta}$ is uniformly distributed between $1$ and $-1$ for an isotropic source. The black photon path also demonstrates why the photons around the line center are not captured in the near-zone SED (Fig.~\ref{fig:spec}): they are farther away than $1h^{-1}~{\rm cMpc}$ from the source on the sky plane. The photons can be observed in the near-zone only when $\theta$ is close to either $0$ or $\pi$ radians, and those are the ones whose the emergent wavelength is either $v_{\alpha}\approx v_{\alpha,i}$ or $-v_{\alpha,i}$.

The right panel of Figure~\ref{fig:LSP} describes the $v_{\alpha,i}= 200~{\rm km}~{\rm s}^{-1}$ case, which represents the photons emitted on the red side of Ly$\alpha$ close to the resonant scattering limit ($v_{\alpha,i}=\sqrt{GM_h/r_s}$). In this case, the peculiar infall velocity at the scattering location ($\sim 200~{\rm km}~{\rm s}^{-1}$) is comparable to $v_{\alpha,i}$, and the IGM peculiar motion can significantly affect the wavelength during scattering events. Here, we use a color discontinuity in the arrows to illustrate how the scattering changes the wavelength depending on the initial direction. When a photon initially propagates away from the observer and is scattered toward the observer at the scattering surface (see the blue path in the figure), the infall motion pointing toward the observer blueshifts the photon enough to place it on the blue side in the emergent spectrum. Conversely, a photon initially headed toward the observer would experience a substantial redshift upon scattering (see the red path). The near-zone SED is not much weaker than the total SED in this case because most of the photons are scattered in the near-zone of the galaxy due to the relatively small scattering surface.

In the case of a small HII bubble (galaxy \#0504), the scattering process within the HII region is similar, but a significant fraction of the photons go through additional scatterings in the surrounding HI region. These scattering events increase the photon path and redshift the photons further, suppressing the blue-side SED and instead enhancing the red-side SED. Also, this redward SED is not captured in the near-zone SED because the scattering makes the photons more extended in space. This explains the difference in the SED between the cases of large and small HII regions (Fig.~\ref{fig:spec}).

\subsection{Spectrum of Scattered Light: Realistic Source Cases} \label{sec:spectrum2}

Next, we consider more realistic cases where the source SEDs have extended profiles. The emergent scattered light SED for an arbitrary emission profile, $F(v_\alpha)$, can be obtained by superposing the monochromatic source cases:
\begin{equation} \label{eq:weight}
F(v_\alpha) = \int^{v_{\alpha,i}^{\rm max}}_{v_{\alpha,i}^{\rm min}} F_m(v_\alpha|v_{\alpha,i})W(v_{\alpha,i}) dv_{\alpha,i},
\end{equation}
where $F_m(v_\alpha|v_{\alpha,i})$ is the emergent SED of the monochromatic source emitting at $v_{\alpha,i}$, and the weight function $W(v_{\alpha,i})$ is given by the intrinsic emission SED of the source galaxy. For each photon, we draw $v_{\alpha,i}$ from a uniform distribution between $v_{\alpha,i}^{\rm max}=1000~{\rm km}~{\rm s}^{-1}$ and $v_{\alpha,i}^{\rm min}=-1000~{\rm km}~{\rm s}^{-1}$ to cover the extent of the typical Ly$\alpha$ emission spectrum of star-forming galaxies. Then, we calculate scattered light SED for the intrinsic emission model of our choice by weighting each photon by the intrinsic emission profile, $W(v_{\alpha,i})$, in the probability distribution. 

\begin{figure*}
\begin{center}
\includegraphics[width=0.35\paperwidth]{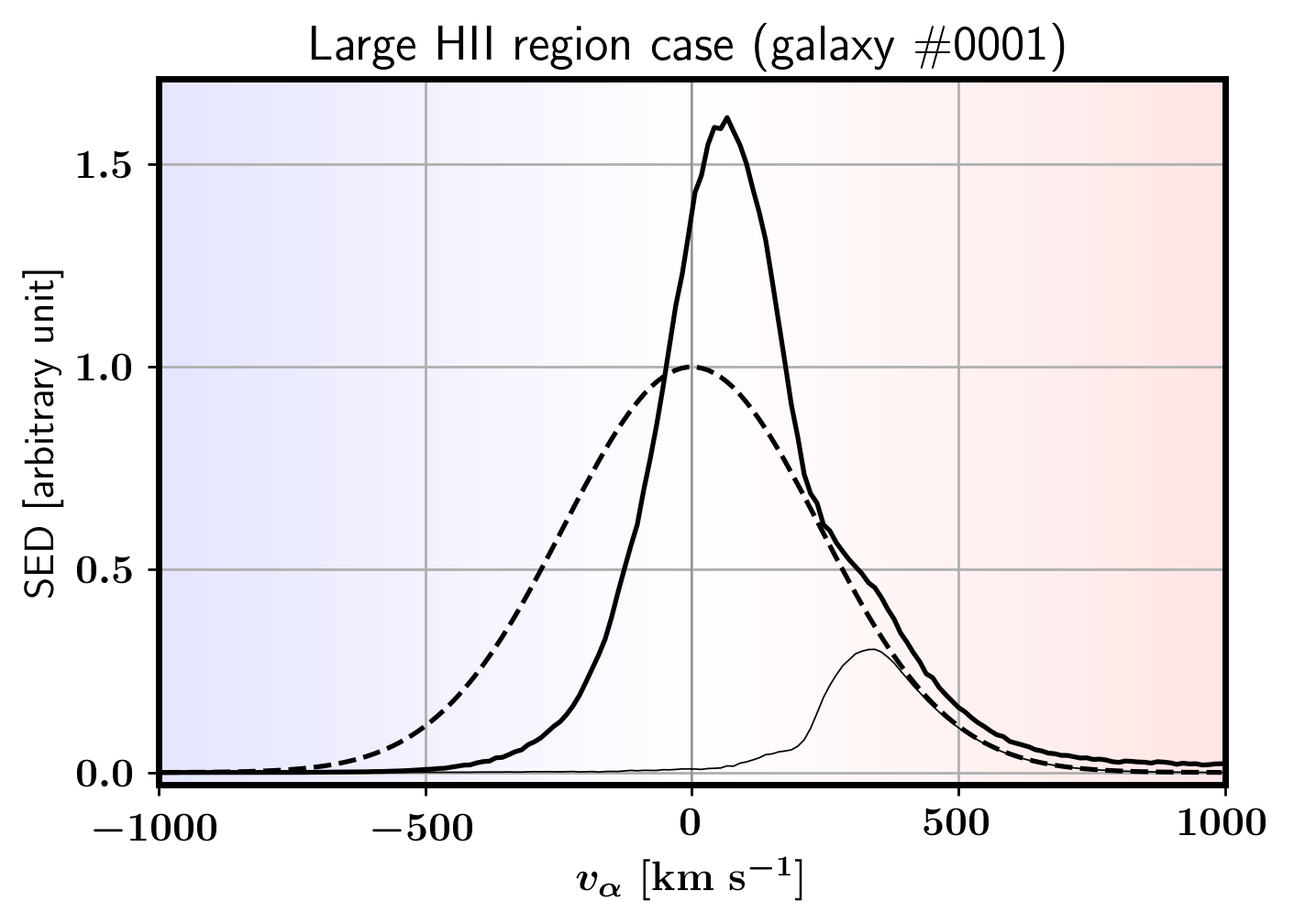}
\includegraphics[width=0.35\paperwidth]{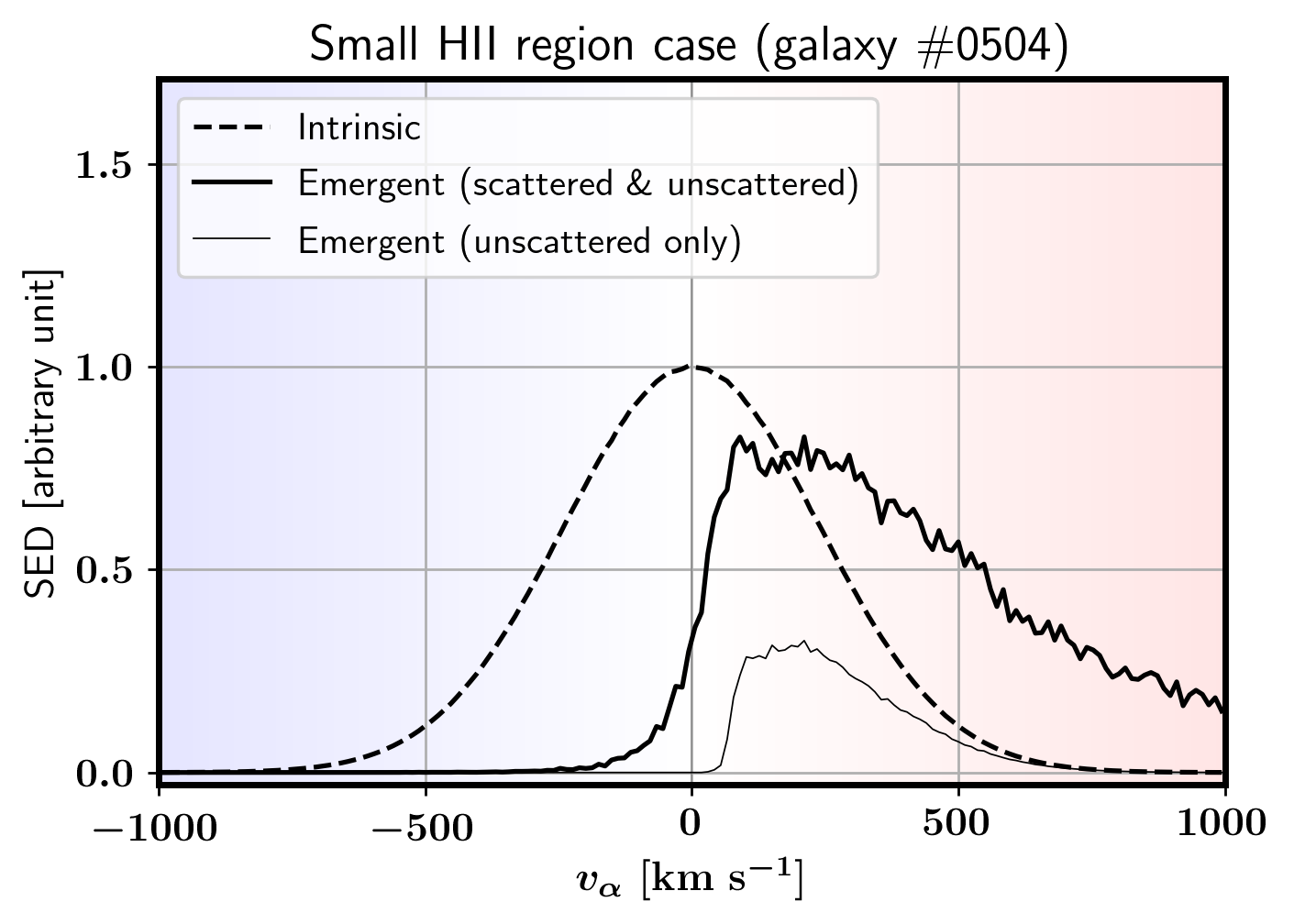}
\caption{\label{fig:CP} The intrinsic SED of the central peak emission model (dashed), the emergent SED of the unscattered light directly coming from the source (thin solid), and the emergent SED of both scattered and unscattered light (thick solid). The left panel is for the photons emitted by galaxy \#0001, and the right panel is for those emitted by galaxy \#0504. }
\end{center}
\end{figure*}

In this work, we consider three models for the intrinsic emission. One model has a broad Gaussian peak centered at the resonance with its full width at FWHM of $V_1=800~{\rm km}~{\rm s}^{-1}$:
\begin{equation}
W_{\rm CP}(v_{\alpha,i})=\exp\left(-\left[\frac{v_{\alpha,i}}{V_1/2.355}\right]^2\right).
\end{equation}
Here, the factor 2.355 is to make $V_1$ the FWHM of the profile. The other two models are the red- and double-peak models, where we place an off-center Gaussian peak either on the red side only or on both sides. The weight functions  are
\begin{equation} \label{eq:RP}
W_{\rm RP}(v_{\alpha,i})=\exp\left(-\left[\frac{v_{\alpha,i}-V_2}{V_2/2.355}\right]^2\right)
\end{equation}
and
\begin{equation}\label{eq:DP}
\begin{aligned}
&W_{\rm DP}(v_{\alpha,i})=\\
&\exp\left(-\left[\frac{v_{\alpha,i}-V_2}{V_2/2.355}\right]^2\right)  +\exp\left(-\left[\frac{v_{\alpha,i}+V_2}{V_2/2.355}\right]^2\right),
\end{aligned}
\end{equation}
where we assume $V_2=300~{\rm km}~{\rm s}^{-1}$ for the offset and the FWHM of the peaks.

The central peaks model is similar to the results from recent galaxy-scale radiative transfer simulation studies \citep[e.g.,][]{2021arXiv211113721S}. In those simulations, star-forming clouds initially radiate at the Ly$\alpha$ resonance, and the profile is broadened due to the turbulent and rotational motion of the interstellar medium (ISM) within the source galaxy. The red-peak model is supported by observation at $z\lesssim 3$, where the IGM is considered to be transparent to Ly$\alpha$ photons. The outflows in the circumgalactic medium are known to suppress the blue-side emission from the star-forming ISM \citep[e.g.,][]{2016ApJ...820..130Y}. The double-peak model is motivated by recent simulation studies suggesting that high-$z$ galaxies may have more porous ISM due to stronger star-formation feedback, allowing more blue-side photons to escape the galaxy. The real shape of the intrinsic emission profile is not well constrained today and needs further studies.

We will use the central peak model to demonstrate how the scattered light is processed in the IGM. Then, we compare the results of the red-peak and double-peak models to assess the impact of the blueward emission on the scattered light, which cannot be seen from the direct observation of the unscattered light. 

\subsubsection{Central Peak Model}

We show the intrinsic emission SED, the unscattered light SED, and the SED of both unscattered and scattered light for galaxies \#0001 and \#0504 in Figure~\ref{fig:CP}. We show the total SED sampled from the entire volume that we calculated Ly$\alpha$ RT ($r_\perp<8h^{-1}~{\rm Mpc}$). 

Comparing the unscattered light SED to the intrinsic SED shows that the photons with $v_{\alpha,i}\lesssim250~(150)~{\rm km}~{\rm s}^{-1}$ are completely scattered by the IGM in the case of a large (small) HII region. As detailed in P21, this truncation wavelength is set by the circular velocity of the halo, $V_c=\sqrt{GM_h/r_{200}}$. Above the circular velocity ($v_\alpha>V_c$), the unscattered SED converges to the intrinsic SED in the case of a large HII region, but it still remains significantly lower in the small HII region because the damping-wing opacity of the HI region scatters a fraction of these photons on the red side.  

Since we do not consider any absorption by dust in the IGM, the integrated SED of the intrinsic emission is the same as that of the scattered and unscattered light combined. Comparing the two cases shows that the scattered light generally ends up redder than it was at emission. As we described in the monochromatic cases, this is due to the scattering event increasing the path length for the scattered light. In the case of a small HII region, the scattered light adds more to the red side due to the additional scatterings in the HI region.

\subsubsection{Red- and Double-Peak Models} \label{sec:red_double_peak_model}

\begin{figure*}
\begin{center}
\includegraphics[width=0.38\paperwidth]{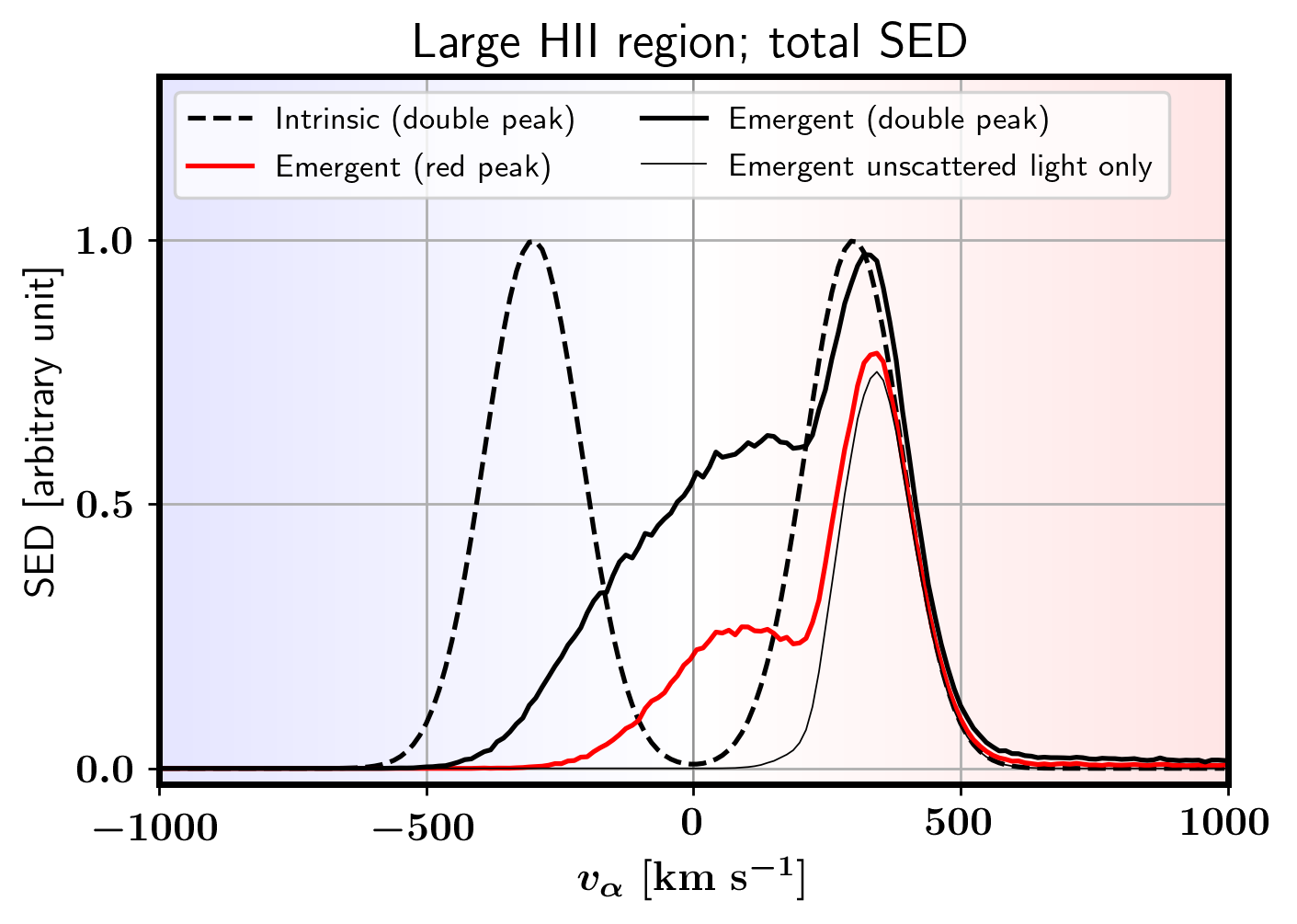}
\includegraphics[width=0.38\paperwidth]{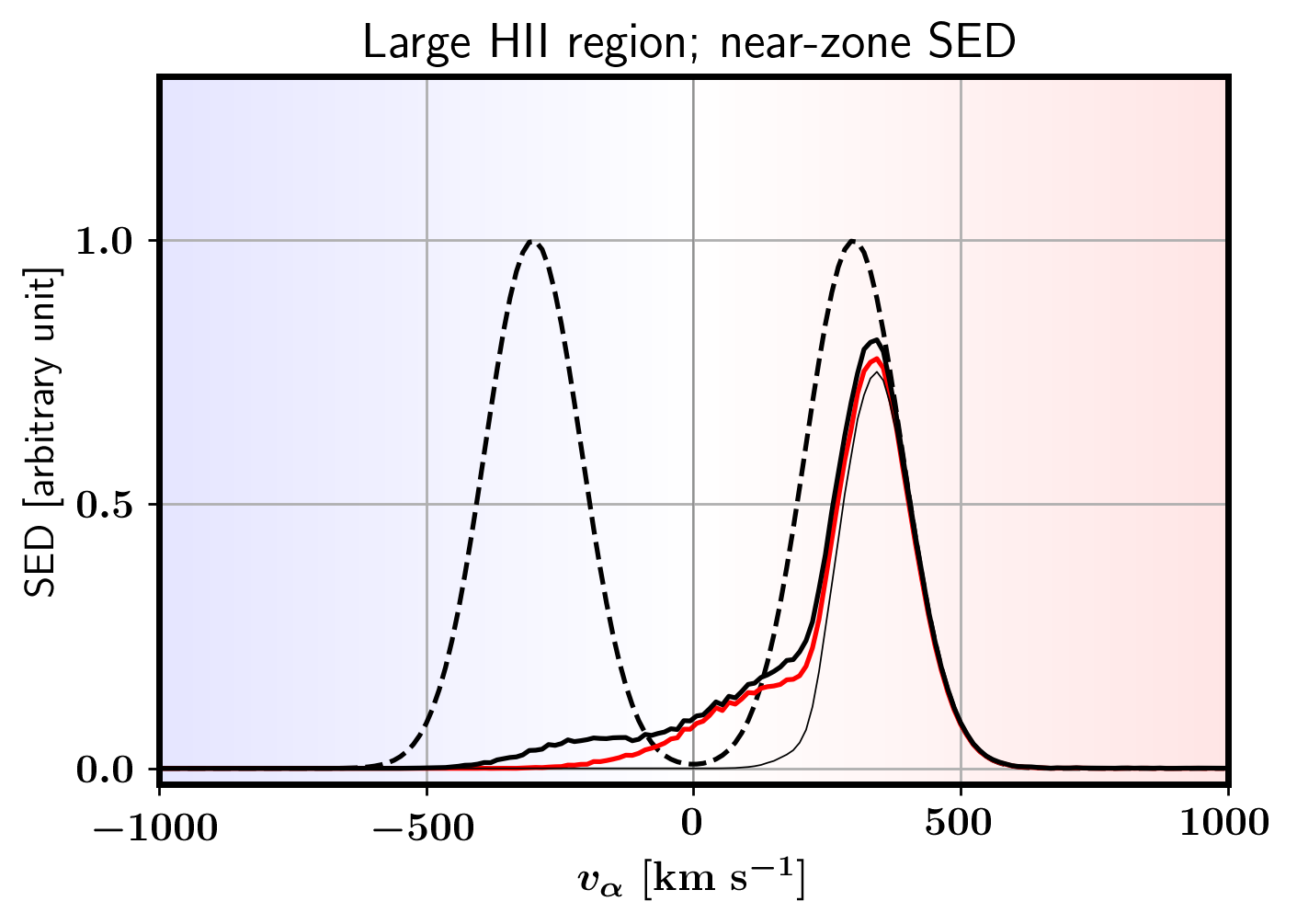}
\includegraphics[width=0.38\paperwidth]{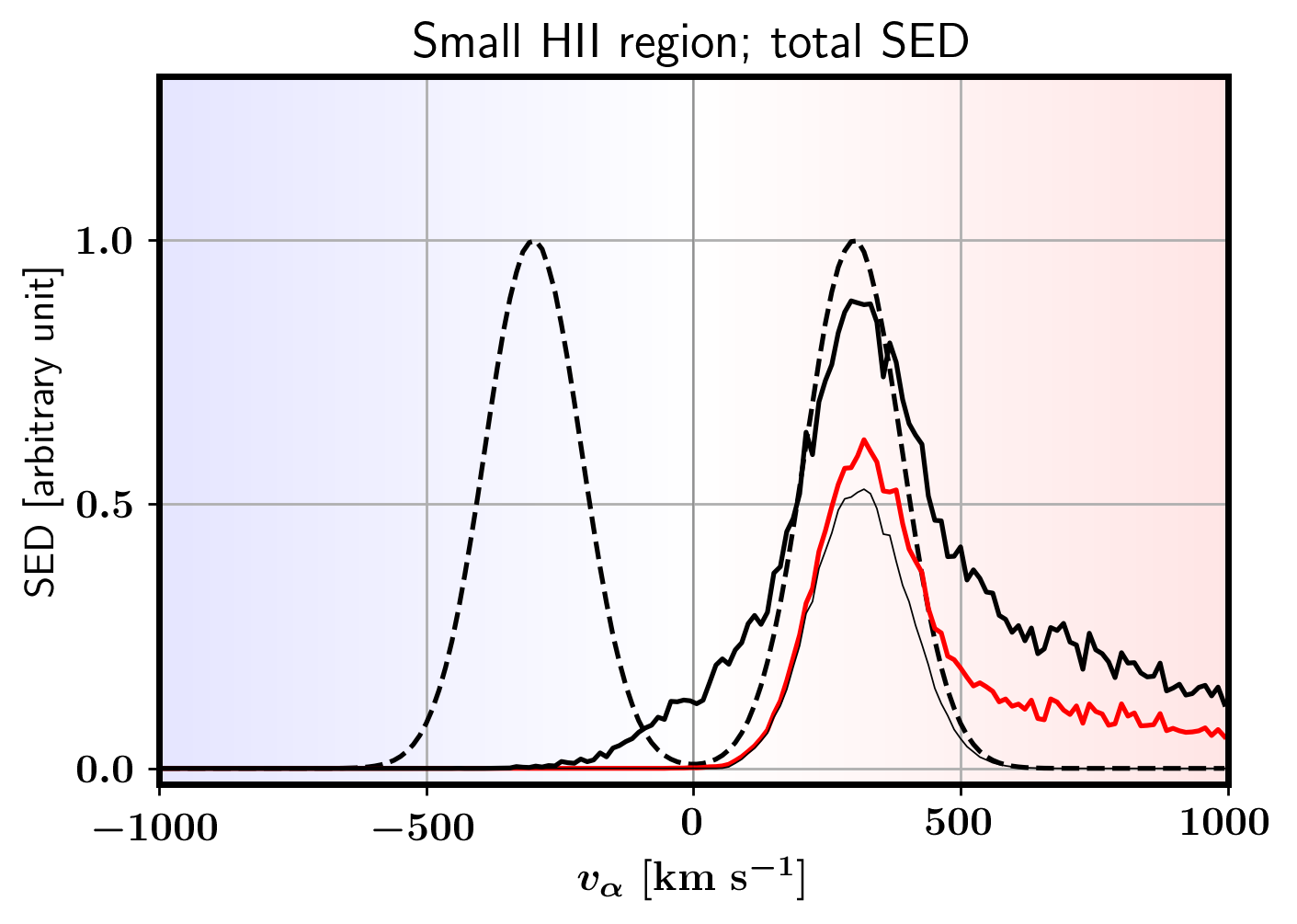}
\includegraphics[width=0.38\paperwidth]{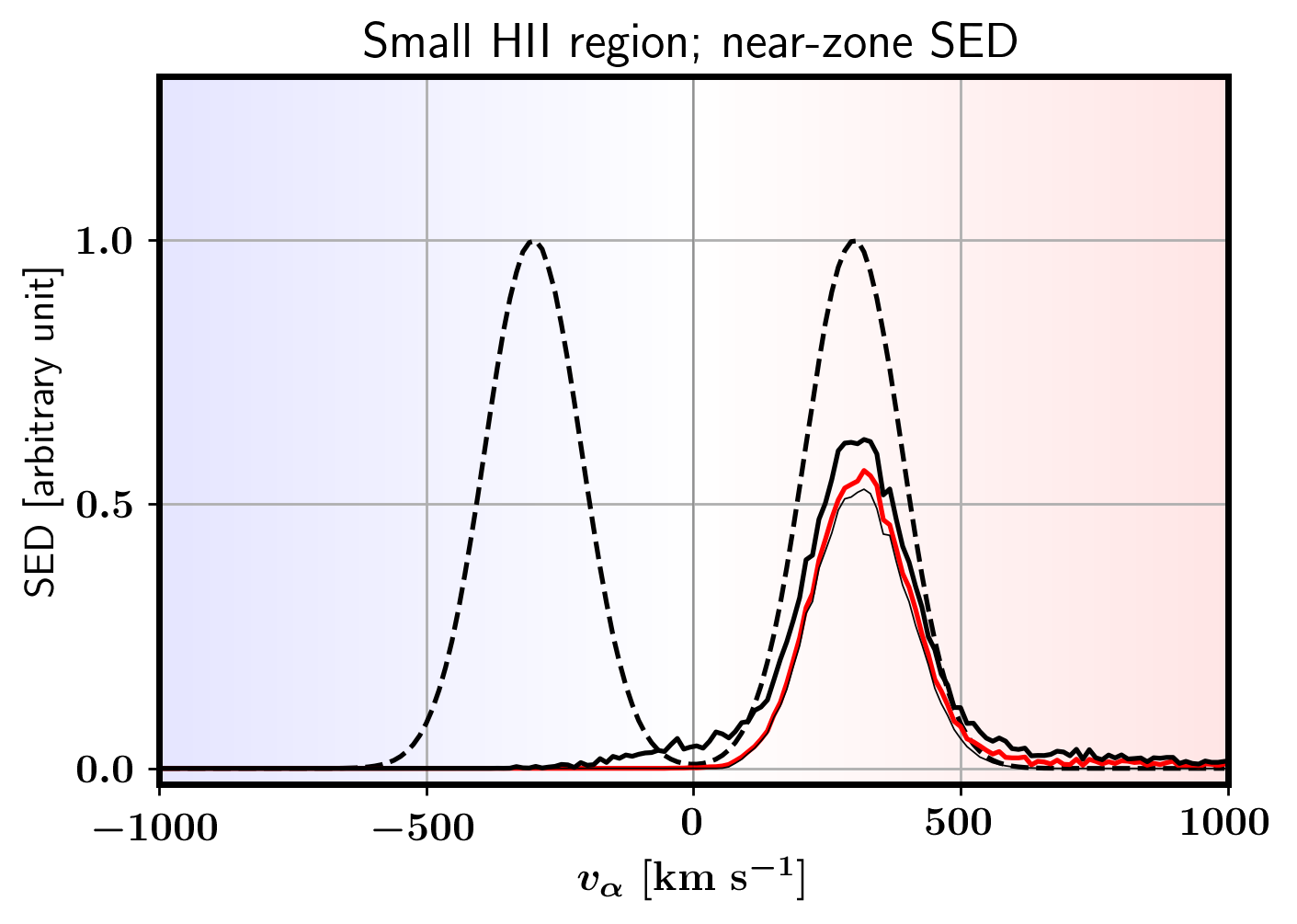}
\caption{\label{fig:dvap2} The SEDs of scattered light for the red-peak and double-peak models are shown as the red and black solid lines, respectively. The SED of the unscattered light is shown as the thin black solid line, and that of the intrinsic emission is shown as the gray dashed line. The upper panels are for photons emitted from galaxy \#0001, and the lower ones are for those from galaxy \#0504. The left and right panels show the results for all the sampled photons and the photons within $8h^{-1}$ and $1h^{-1}~{\rm Mpc}$ or 4.5 and 0.6 arcminutes from the source on the sky, respectively.}
\end{center}
\end{figure*}

In Figure~\ref{fig:dvap2}, we compare the scattered light in the red-peak and double-peak source models for the cases of large and small HII regions for total and near-zone SEDs. We note that the unscattered light SED is the same in both models because the blue-side emission is completely scattered by the IGM on its way and does not appear in the unscattered light SED.  We also show the intrinsic emission profile of the double-peak model. For the red-peak model, we take the red-side peak of the double-peak model as described by Equations~(\ref{eq:RP}) and (\ref{eq:DP}). 

In the large HII region (upper panels), the scattered light adds a blueward tail to the combined SED while the unscattered light is truncated near the circular velocity of the halo ($v_\alpha\sim 250~{\rm km}~{\rm s}^{-1}$). In the total SED case (upper left panel), the blue peak makes a significant difference in the SED: the scattered light from the double-peak case adds a thicker and more extended tail (down to $\sim-500~{\rm km}~{\rm s}^{-1}$) than the red-peak case does ($\sim-250~{\rm km}~{\rm s}^{-1}$). However, the blue-peak contribution is much weaker in the near-zone SED (upper right panel), because the blue-side emission forms more extended scattered light in the sky, as we observed in the monochromatic cases. We also repeat this calculation with nonradial emission at $r_{200}$ in the appendix to confirm that the results do not depend sensitively on the initial photon direction at emission.

In the small HII region (lower panels), the scattered light tends to be distributed over a wider wavelength range due to the additional scatterings in the HI region resulting in more redshifting of the scattered light. The scattered light adds an extended redward tail in the SED well beyond $v_\alpha=1000~{\rm km}~{\rm s}^{-1}$, as in the monochromatic cases and in the central peak model. The blueward tail is also present, but it appears much weaker than in the case of a large HII region. Also, the contribution from scattered light in the near-zone SED (lower right panel) is smaller than in the large HII region case for both models, because the scatterings in the HI region spread both red- and blue-side photons to a larger patch of the sky.

\section{Summary and Discussion}

We have developed a Monte Carlo Ly$\alpha$ RT simulation code to trace the Ly$\alpha$ photons emitted from high-$z$ galaxies in the intergalactic medium. The code can run on an arbitrary three-dimensional mesh of density, velocity, ionization, and temperature. We do not use any acceleration schemes often adopted in similar works. 

We have tested the code for several problems with analytic solutions, including a monochromatic source in a static uniform isothermal slab \citep{1973MNRAS.162...43H,1990ApJ...350..216N} and sphere \citep{2006ApJ...649...14D}. We also test the code against the results from already published works for simple geometry and kinematics, where a sphere of gas has Hubble-like radially outward (or inward) motion \citep{2002ApJ...578...33Z,2006ApJ...645..792T,2006ApJ...649...14D,2009ApJ...696..853L}. The emergent spectra and their dependence on the physical parameters agree with the results from the previous works, indicating that our code is reliable.

We ran our code for volumes near two galaxies in the $z=7$ snapshot of the CoDaII simulation to explore the physics of Ly$\alpha$ photon scattering during reionization. Based on the results, we explained how the scattering location is determined by the initial wavelength and cosmological redshift in the HII regions, and how the damping-wing opacity in the HI region can affect the scattering locations. Then, we explained how the geometry of the scattering location is related to the emergent SED of the scattered light. We also present the results for the near-zone of the galaxy ($0.56~{\rm arcmin}$ from source) in comparison to the total SED within the extent of the simulation ($4.5~{\rm arcmin}$ from source).

In an HII region, photons propagate freely until they redshift to the Ly$\alpha$ resonance and are scattered by residual neutral hydrogen atoms. The scattering events can increase the frequency by upscattering the photons with the peculiar motion of the gas, and also decrease the frequency due to extra cosmological redshift from the increased path length to the observer. Typically, the photons emitted on the blue side of the resonance experience more redshifting because their path length is more dramatically increased by scattering events. Those emitted on the red side, in contrast, face stronger gravitational infall motion of the IGM, giving a larger boost to their frequencies. The blue-side emission makes a significant difference in the emergent spectrum of the scattered light, which potentially allows us to discriminate different intrinsic emission profiles on the blue side, which is impossible from the unscattered light. However, the difference is much smaller if we limit the light collection to the near-zone $(r_\perp < 0.6~{\rm arcmin})$ of the galaxy because the blue-side emission ends up more diffuse and extended in the sky. These findings are 
broadly consistent with what was reported by \citet[][See Sec. 4 of their work]{2010ApJ...716..574Z}.

The ionization state of the IGM is another crucial factor. If the surrounding HII region is small ($\lesssim 2h^{-1}~{\rm cMpc}$), the damp-wing opacity of the nearby HI region becomes significant even for the photons on the red side of the resonance. The photons go through more scattering events in the HI region, resulting in much more spread in both frequency and space. This suggests that the detectability of the scattered light would steeply drop toward high $z$ as the HII bubble are expected to be smaller at earlier times. 

Our results provide a theoretical framework for interpreting future observations to constrain the properties of the source galaxies. In realistic observations with IFUs, there will be multiple neighboring galaxies within a field of several square arcminutes around the target galaxy, and the light from the neighboring galaxies would mix with that from the target, making the interpretation non-trivial. The collective spatial intensity map of scattered Ly$\alpha$ photons can be studied statistically, e.g. through the intensity power spectrum that could also probe the physical state of the IGM \citep{2018ApJ...863L...6V}. Our study, focused on individual objects, would still be applicable to very bright objects whose scattered Ly$\alpha$ intensity stands out against diffuse background. Further studies may be needed depending on the specifics of interested surveys. 

We note that we treat the ISM and CGM as a black box in this work when simulating the photons from the virial radius of the galaxy. Understanding the intrinsic emission exiting the CGM requires dedicated small-scale simulations and is under active investigation by other numerical studies \citep[e.g.][]{2021arXiv211113721S}. Our results can flexibly accommodate any intrinsic profile from other studies and produce the corresponding scattered light SED using Equation~(\ref{eq:weight}).

In future work, we plan to extend our analysis to the surface brightness of the scattered light, which is relevant to future intensity mapping surveys such as SPHEREx.  We shall also explore the dependence on the viewing angle of the scattered light to address the possible variation in observational constraints.


\section*{Acknowledgements}
We thank the anonymous referee and A. Smith for helpful comments on this paper. H.P. was supported by the World Premier International Research Center Initiative (WPI), MEXT, Japan and JSPS KAKENHI grant No. 19K23455. The numerical computations of this work were carried out on the gfarm computing cluster of the Kavli Institute for Physics and Mathematics of the Universe and a high-performance computing cluster at the Korea Astronomy and Space Science Institute. H.S. was supported by the Basic Science Research Program through the National Research Foundation of Korea (NRF) funded by the Ministry of Education (2020R1I1A1A01069228).  K.A. was supported by NRF-2016R1D1A1B04935414, 2021R1A2C1095136 and 2016R1A5A1013277.  P.R.S. was supported in part by US NSF grant AST-1009799, NASA grant NNX11AE09G, NASA/JPL grant RSA Nos. 1492788 and 1515294, and supercomputer resources from NSF XSEDE grant TG-AST090005 and the Texas Advanced Computing Center (TACC) at the University of Texas at Austin. T.D. was supported by the National Science Foundation Graduate Research Fellowship Program under grant No. DGE-1610403. J.S. acknowledges support from the ANR LOCALIZATION project, grant ANR-21-CE31-0019 of the French Agence Nationale de la Recherche. I.T.I. was supported by the Science and Technology Facilities Council (grant Nos. ST/I000976/1 and ST/T000473/1) and the Southeast Physics Network (SEPNet). I.J. acknowledges support from NASA under award number 80GSFC21M0002.

\appendix 

\begin{figure*}
\begin{center}
\includegraphics[width=0.38\paperwidth]{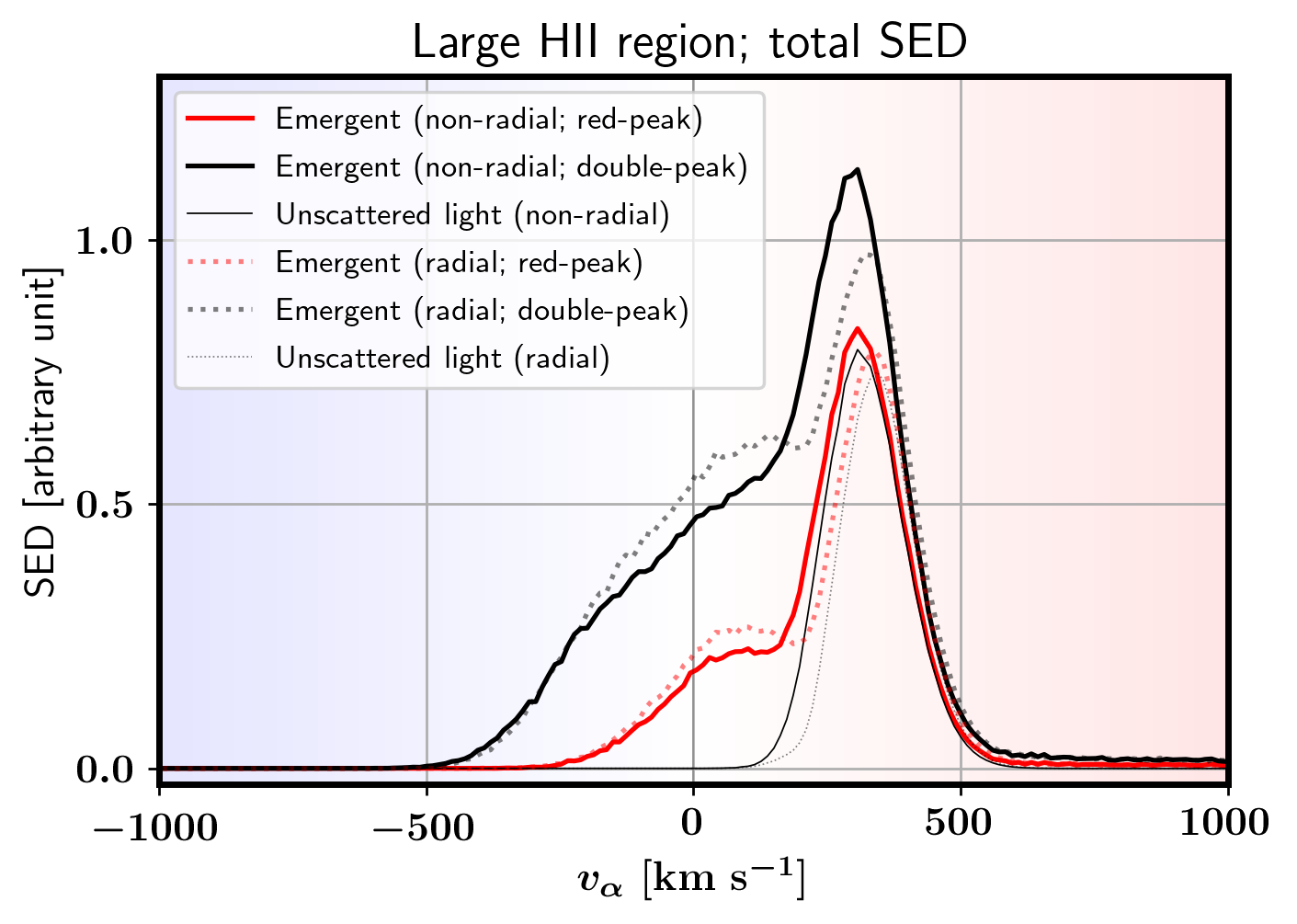}
\includegraphics[width=0.38\paperwidth]{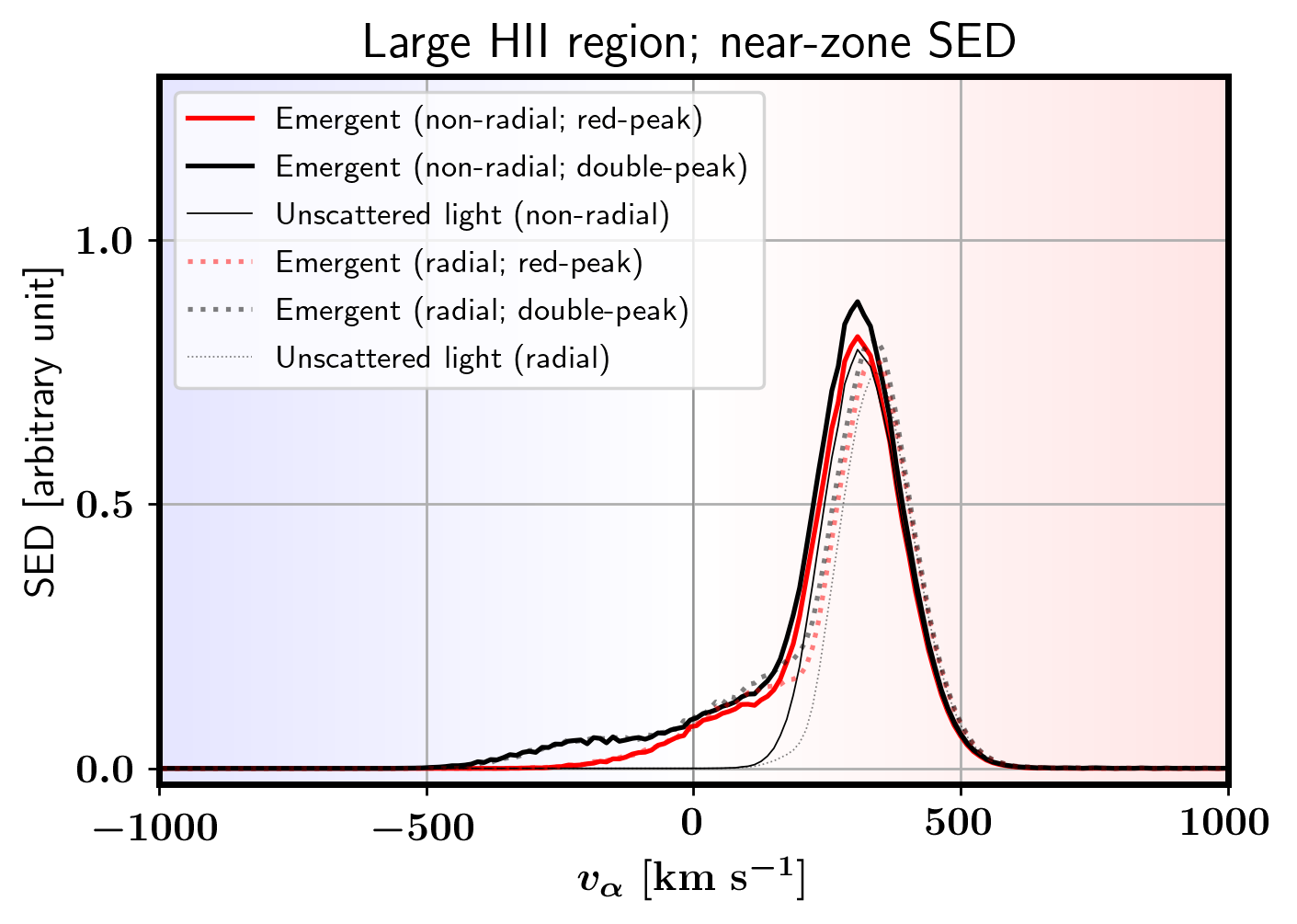}
\caption{\label{fig:dvap2_nre} The total (left) and near-zone (right) SEDs for nonradial emission. Similarly to in Figure~\ref{fig:dvap2}, we show the results for red-peak and double-peak models by the red and black thick solid lines, and the case of unscattered light only is shown by the thin black line. The dotted lines are from the radial emission cases of Figure~\ref{fig:dvap2}. }
\end{center}
\end{figure*}

\section{Dependence on photon direction at emission}
In this work, we assume all the Ly$\alpha$ photons emitted at $r_{200}$ are in the radial direction, but this is not true in reality, as some photons would be scattered within the CGM and change their directions. Given that $r_{200}$ is small compared to the distance to the first scattering location of most Ly$\alpha$ photons, we do not expect the results to depend sensitively on this assumption. To test how the results depend on the initial photon direction, we run our Ly$\alpha$ RT for a case of nonradial emission, in which we randomly draw the initial photon direction from the hemisphere pointing toward the radial direction. Thus, the actual distribution of the emission direction would lie somewhere between the perfectly radial case and this semi-isotropic (nonradial) case that we test here. 

We repeat our calculation with the nonradial emission for the case pf a large HII bubble of galaxy \#0001 and show the results in Figure~\ref{fig:dvap2_nre}. For comparison, we show the results from the case of radial emission (solid lines in the upper panels of Fig.~\ref{fig:dvap2}) as a dotted line of the same color and line thickness. Both the total and near-zone SEDs show only a small difference between the two emission cases, confirming that the initial photon direction is a minor factor in the those results.

\bibliographystyle{aasjournal}
\bibliography{cosmo_new}

\begin{thebibliography}{}
\expandafter\ifx\csname natexlab\endcsname\relax\def\natexlab#1{#1}\fi
\providecommand{\url}[1]{\href{#1}{#1}}
\providecommand{\dodoi}[1]{doi:~\href{http://doi.org/#1}{\nolinkurl{#1}}}
\providecommand{\doeprint}[1]{\href{http://ascl.net/#1}{\nolinkurl{http://ascl.net/#1}}}
\providecommand{\doarXiv}[1]{\href{https://arxiv.org/abs/#1}{\nolinkurl{https://arxiv.org/abs/#1}}}

\bibitem[{{Adams}(1971)}]{1971ApJ...168..575A}
{Adams}, T.~F. 1971, \apj, 168, 575, \dodoi{10.1086/151111}

\bibitem[{{Ahn} {et~al.}(2000){Ahn}, {Lee}, \& {Lee}}]{2000JKAS...33...29A}
{Ahn}, S.-H., {Lee}, H.-W., \& {Lee}, H.~M. 2000, Journal of Korean
  Astronomical Society, 33, 29

\bibitem[{{Ahn} {et~al.}(2001){Ahn}, {Lee}, \& {Lee}}]{2001ApJ...554..604A}
---. 2001, \apj, 554, 604, \dodoi{10.1086/321374}

\bibitem[{{Ahn} {et~al.}(2002){Ahn}, {Lee}, \& {Lee}}]{2002ApJ...567..922A}
---. 2002, \apj, 567, 922, \dodoi{10.1086/338497}

\bibitem[{{Bunker} {et~al.}(2003){Bunker}, {Smith}, {Spinrad}, {Stern}, \&
  {Warren}}]{Bunker2003}
{Bunker}, A., {Smith}, J., {Spinrad}, H., {Stern}, D., \& {Warren}, S. 2003,
  \apss, 284, 357, \dodoi{10.1023/A:1024038312479}

\bibitem[{{Camps} {et~al.}(2021){Camps}, {Behrens}, {Baes}, {Kapoor}, \&
  {Grand}}]{2021ApJ...916...39C}
{Camps}, P., {Behrens}, C., {Baes}, M., {Kapoor}, A.~U., \& {Grand}, R. 2021,
  \apj, 916, 39, \dodoi{10.3847/1538-4357/ac06cb}

\bibitem[{{Cantalupo} {et~al.}(2012){Cantalupo}, {Lilly}, \&
  {Haehnelt}}]{2012MNRAS.425.1992C}
{Cantalupo}, S., {Lilly}, S.~J., \& {Haehnelt}, M.~G. 2012, \mnras, 425, 1992,
  \dodoi{10.1111/j.1365-2966.2012.21529.x}

\bibitem[{{Cantalupo} {et~al.}(2005){Cantalupo}, {Porciani}, {Lilly}, \&
  {Miniati}}]{2005ApJ...628...61C}
{Cantalupo}, S., {Porciani}, C., {Lilly}, S.~J., \& {Miniati}, F. 2005, \apj,
  628, 61, \dodoi{10.1086/430758}

\bibitem[{{Castellano} {et~al.}(2018){Castellano}, {Pentericci}, {Vanzella},
  {Marchi}, {Fontana}, {Dayal}, {Ferrara}, {Hutter}, {Carniani}, {Cristiani},
  {Dickinson}, {Gallerani}, {Giallongo}, {Giavalisco}, {Grazian}, {Maiolino},
  {Merlin}, {Paris}, {Pilo}, \& {Santini}}]{2018ApJ...863L...3C}
{Castellano}, M., {Pentericci}, L., {Vanzella}, E., {et~al.} 2018, \apjl, 863,
  L3, \dodoi{10.3847/2041-8213/aad59b}

\bibitem[{{Colbert} {et~al.}(2011){Colbert}, {Scarlata}, {Teplitz}, {Francis},
  {Palunas}, {Williger}, \& {Woodgate}}]{Colbert2011}
{Colbert}, J.~W., {Scarlata}, C., {Teplitz}, H., {et~al.} 2011, \apj, 728, 59,
  \dodoi{10.1088/0004-637X/728/1/59}

\bibitem[{{Croft} {et~al.}(2018){Croft}, {Miralda-Escud{\'e}}, {Zheng},
  {Blomqvist}, \& {Pieri}}]{2018MNRAS.481.1320C}
{Croft}, R. A.~C., {Miralda-Escud{\'e}}, J., {Zheng}, Z., {Blomqvist}, M., \&
  {Pieri}, M. 2018, \mnras, 481, 1320, \dodoi{10.1093/mnras/sty2302}

\bibitem[{{Curtis-Lake} {et~al.}(2012){Curtis-Lake}, {McLure}, {Pearce},
  {Dunlop}, {Cirasuolo}, {Stark}, {Almaini}, {Bradshaw}, {Chuter}, {Foucaud},
  \& {Hartley}}]{2012MNRAS.422.1425C}
{Curtis-Lake}, E., {McLure}, R.~J., {Pearce}, H.~J., {et~al.} 2012, \mnras,
  422, 1425, \dodoi{10.1111/j.1365-2966.2012.20720.x}

\bibitem[{{Dijkstra}(2014)}]{2014PASA...31...40D}
{Dijkstra}, M. 2014, \pasa, 31, e040, \dodoi{10.1017/pasa.2014.33}

\bibitem[{{Dijkstra}(2017)}]{2017arXiv170403416D}
---. 2017, arXiv e-prints, arXiv:1704.03416.
\newblock \doarXiv{1704.03416}

\bibitem[{{Dijkstra} {et~al.}(2006){Dijkstra}, {Haiman}, \&
  {Spaans}}]{2006ApJ...649...14D}
{Dijkstra}, M., {Haiman}, Z., \& {Spaans}, M. 2006, \apj, 649, 14,
  \dodoi{10.1086/506243}

\bibitem[{{Dijkstra} \& {Loeb}(2009)}]{Dijkstra2009}
{Dijkstra}, M., \& {Loeb}, A. 2009, \mnras, 400, 1109,
  \dodoi{10.1111/j.1365-2966.2009.15533.x}

\bibitem[{{Endsley} {et~al.}(2021){Endsley}, {Stark}, {Charlot}, {Chevallard},
  {Robertson}, {Bouwens}, \& {Stefanon}}]{2021MNRAS.502.6044E}
{Endsley}, R., {Stark}, D.~P., {Charlot}, S., {et~al.} 2021, \mnras, 502, 6044,
  \dodoi{10.1093/mnras/stab432}

\bibitem[{{Faucher-Gigu{\`e}re} {et~al.}(2010){Faucher-Gigu{\`e}re}, {Kere{\v
  s}}, {Dijkstra}, {Hernquist}, \& {Zaldarriaga}}]{Faucher-Giguere2010}
{Faucher-Gigu{\`e}re}, C.-A., {Kere{\v s}}, D., {Dijkstra}, M., {Hernquist},
  L., \& {Zaldarriaga}, M. 2010, \apj, 725, 633,
  \dodoi{10.1088/0004-637X/725/1/633}

\bibitem[{{Fernandez} \& {Komatsu}(2006)}]{Fernandez2006}
{Fernandez}, E.~R., \& {Komatsu}, E. 2006, \apj, 646, 703,
  \dodoi{10.1086/505126}

\bibitem[{{Fontana} {et~al.}(2010){Fontana}, {Vanzella}, {Pentericci},
  {Castellano}, {Giavalisco}, {Grazian}, {Boutsia}, {Cristiani}, {Dickinson},
  {Giallongo}, {Maiolino}, {Moorwood}, \& {Santini}}]{Fontana2010}
{Fontana}, A., {Vanzella}, E., {Pentericci}, L., {et~al.} 2010, \apjl, 725,
  L205, \dodoi{10.1088/2041-8205/725/2/L205}

\bibitem[{{Francis} {et~al.}(1996){Francis}, {Woodgate}, {Warren}, {M{\o}ller},
  {Mazzolini}, {Bunker}, {Lowenthal}, {Williams}, {Minezaki}, {Kobayashi}, \&
  {Yoshii}}]{Francis1996}
{Francis}, P.~J., {Woodgate}, B.~E., {Warren}, S.~J., {et~al.} 1996, \apj, 457,
  490, \dodoi{10.1086/176747}

\bibitem[{{Gronke} {et~al.}(2021){Gronke}, {Ocvirk}, {Mason}, {Matthee},
  {Bosman}, {Sorce}, {Lewis}, {Ahn}, {Aubert}, {Dawoodbhoy}, {Iliev},
  {Shapiro}, \& {Yepes}}]{2021MNRAS.508.3697G}
{Gronke}, M., {Ocvirk}, P., {Mason}, C., {et~al.} 2021, \mnras, 508, 3697,
  \dodoi{10.1093/mnras/stab2762}

\bibitem[{{Haiman} {et~al.}(2000){Haiman}, {Spaans}, \&
  {Quataert}}]{Haiman2000}
{Haiman}, Z., {Spaans}, M., \& {Quataert}, E. 2000, \apjl, 537, L5,
  \dodoi{10.1086/312754}

\bibitem[{{Hansen} \& {Oh}(2006)}]{2006MNRAS.367..979H}
{Hansen}, M., \& {Oh}, S.~P. 2006, \mnras, 367, 979,
  \dodoi{10.1111/j.1365-2966.2005.09870.x}

\bibitem[{{Harikane} {et~al.}(2019){Harikane}, {Ouchi}, {Ono}, {Fujimoto},
  {Donevski}, {Shibuya}, {Faisst}, {Goto}, {Hatsukade}, {Kashikawa}, {Kohno},
  {Hashimoto}, {Higuchi}, {Inoue}, {Lin}, {Martin}, {Overzier}, {Smail},
  {Toshikawa}, {Umehata}, {Ao}, {Chapman}, {Clements}, {Im}, {Jing},
  {Kawaguchi}, {Lee}, {Lee}, {Lin}, {Matsuoka}, {Marinello}, {Nagao},
  {Onodera}, {Toft}, \& {Wang}}]{2019ApJ...883..142H}
{Harikane}, Y., {Ouchi}, M., {Ono}, Y., {et~al.} 2019, \apj, 883, 142,
  \dodoi{10.3847/1538-4357/ab2cd5}

\bibitem[{{Harrington}(1973)}]{1973MNRAS.162...43H}
{Harrington}, J.~P. 1973, \mnras, 162, 43, \dodoi{10.1093/mnras/162.1.43}

\bibitem[{{Higuchi} {et~al.}(2019){Higuchi}, {Ouchi}, {Ono}, {Shibuya},
  {Toshikawa}, {Harikane}, {Kojima}, {Chiang}, {Egami}, {Kashikawa},
  {Overzier}, {Konno}, {Inoue}, {Hasegawa}, {Fujimoto}, {Goto}, {Ishikawa},
  {Ito}, {Komiyama}, \& {Tanaka}}]{2019ApJ...879...28H}
{Higuchi}, R., {Ouchi}, M., {Ono}, Y., {et~al.} 2019, \apj, 879, 28,
  \dodoi{10.3847/1538-4357/ab2192}

\bibitem[{{Hu} {et~al.}(2021){Hu}, {Wang}, {Infante}, {Rhoads}, {Zheng},
  {Yang}, {Malhotra}, {Barrientos}, {Jiang}, {Gonz{\'a}lez-L{\'o}pez},
  {Prieto}, {Perez}, {Hibon}, {Galaz}, {Coughlin}, {Harish}, {Kong}, {Kang},
  {Khostovan}, {Pharo}, {Valdes}, {Wold}, {Walker}, \&
  {Zheng}}]{2021NatAs...5..485H}
{Hu}, W., {Wang}, J., {Infante}, L., {et~al.} 2021, Nature Astronomy, 5, 485,
  \dodoi{10.1038/s41550-020-01291-y}

\bibitem[{{Hutter} {et~al.}(2014){Hutter}, {Dayal}, {Partl}, \&
  {M{\"u}ller}}]{2014MNRAS.441.2861H}
{Hutter}, A., {Dayal}, P., {Partl}, A.~M., \& {M{\"u}ller}, V. 2014, \mnras,
  441, 2861, \dodoi{10.1093/mnras/stu791}

\bibitem[{{Iliev} {et~al.}(2008){Iliev}, {Shapiro}, {McDonald}, {Mellema}, \&
  {Pen}}]{2008MNRAS.391...63I}
{Iliev}, I.~T., {Shapiro}, P.~R., {McDonald}, P., {Mellema}, G., \& {Pen},
  U.-L. 2008, \mnras, 391, 63, \dodoi{10.1111/j.1365-2966.2008.13879.x}

\bibitem[{{Jeeson-Daniel} {et~al.}(2012){Jeeson-Daniel}, {Ciardi}, {Maio},
  {Pierleoni}, {Dijkstra}, \& {Maselli}}]{2012MNRAS.424.2193J}
{Jeeson-Daniel}, A., {Ciardi}, B., {Maio}, U., {et~al.} 2012, \mnras, 424,
  2193, \dodoi{10.1111/j.1365-2966.2012.21378.x}

\bibitem[{{Jung} {et~al.}(2019){Jung}, {Finkelstein}, {Dickinson}, {Hutchison},
  {Larson}, {Papovich}, {Pentericci}, {Song}, {Ferguson}, {Guo}, {Malhotra},
  {Mobasher}, {Rhoads}, {Tilvi}, \& {Wold}}]{2019ApJ...877..146J}
{Jung}, I., {Finkelstein}, S.~L., {Dickinson}, M., {et~al.} 2019, \apj, 877,
  146, \dodoi{10.3847/1538-4357/ab1bde}

\bibitem[{{Jung} {et~al.}(2020){Jung}, {Finkelstein}, {Dickinson}, {Hutchison},
  {Larson}, {Papovich}, {Pentericci}, {Straughn}, {Guo}, {Malhotra}, {Rhoads},
  {Song}, {Tilvi}, \& {Wold}}]{2020ApJ...904..144J}
---. 2020, \apj, 904, 144, \dodoi{10.3847/1538-4357/abbd44}

\bibitem[{{Jung} {et~al.}(2021){Jung}, {Papovich}, {Finkelstein}, {Simons},
  {Estrada-Carpenter}, {Backhaus}, {Cleri}, {Finlator}, {Giavalisco}, {Ji},
  {Matharu}, {Momcheva}, {Straughn}, \& {Trump}}]{2021arXiv211114863J}
{Jung}, I., {Papovich}, C., {Finkelstein}, S.~L., {et~al.} 2021, arXiv
  e-prints, arXiv:2111.14863.
\newblock \doarXiv{2111.14863}

\bibitem[{{Kakuma} {et~al.}(2021){Kakuma}, {Ouchi}, {Harikane}, {Ono}, {Inoue},
  {Komiyama}, {Kusakabe}, {Lee}, {Matsuda}, {Matsuoka}, {Mawatari}, {Momose},
  {Shibuya}, \& {Taniguchi}}]{2021ApJ...916...22K}
{Kakuma}, R., {Ouchi}, M., {Harikane}, Y., {et~al.} 2021, \apj, 916, 22,
  \dodoi{10.3847/1538-4357/ac0725}

\bibitem[{{Katz} {et~al.}(2019){Katz}, {Galligan}, {Kimm}, {Rosdahl},
  {Haehnelt}, {Blaizot}, {Devriendt}, {Slyz}, {Laporte}, \&
  {Ellis}}]{2019MNRAS.487.5902K}
{Katz}, H., {Galligan}, T.~P., {Kimm}, T., {et~al.} 2019, \mnras, 487, 5902,
  \dodoi{10.1093/mnras/stz1672}

\bibitem[{{Kikuchihara} {et~al.}(2021){Kikuchihara}, {Harikane}, {Ouchi},
  {Ono}, {Shibuya}, {Itoh}, {Kakuma}, {Inoue}, {Kusakabe}, {Shimasaku},
  {Momose}, {Sugahara}, {Kikuta}, {Saito}, {Kashikawa}, {Zhang}, \&
  {Lee}}]{2021arXiv210809288K}
{Kikuchihara}, S., {Harikane}, Y., {Ouchi}, M., {et~al.} 2021, arXiv e-prints,
  arXiv:2108.09288.
\newblock \doarXiv{2108.09288}

\bibitem[{{Kim} {et~al.}(2020){Kim}, {Yang}, {Zabludoff}, {Smith}, {Jannuzi},
  {Lee}, {Hwang}, \& {Park}}]{2020ApJ...894...33K}
{Kim}, E., {Yang}, Y., {Zabludoff}, A., {et~al.} 2020, \apj, 894, 33,
  \dodoi{10.3847/1538-4357/ab837f}

\bibitem[{{Laursen} {et~al.}(2009){Laursen}, {Razoumov}, \&
  {Sommer-Larsen}}]{2009ApJ...696..853L}
{Laursen}, P., {Razoumov}, A.~O., \& {Sommer-Larsen}, J. 2009, \apj, 696, 853,
  \dodoi{10.1088/0004-637X/696/1/853}

\bibitem[{{Leclercq} {et~al.}(2017){Leclercq}, {Bacon}, {Wisotzki}, {Mitchell},
  {Garel}, {Verhamme}, {Blaizot}, {Hashimoto}, {Herenz}, {Conseil},
  {Cantalupo}, {Inami}, {Contini}, {Richard}, {Maseda}, {Schaye}, {Marino},
  {Akhlaghi}, {Brinchmann}, \& {Carollo}}]{2017A&A...608A...8L}
{Leclercq}, F., {Bacon}, R., {Wisotzki}, L., {et~al.} 2017, \aap, 608, A8,
  \dodoi{10.1051/0004-6361/201731480}

\bibitem[{{Loeb} \& {Rybicki}(1999)}]{1999ApJ...524..527L}
{Loeb}, A., \& {Rybicki}, G.~B. 1999, \apj, 524, 527, \dodoi{10.1086/307844}

\bibitem[{{Malhotra} \& {Rhoads}(2004)}]{2004ApJ...617L...5M}
{Malhotra}, S., \& {Rhoads}, J.~E. 2004, \apjl, 617, L5, \dodoi{10.1086/427182}

\bibitem[{{Mallery} {et~al.}(2012){Mallery}, {Mobasher}, {Capak}, {Kakazu},
  {Masters}, {Ilbert}, {Hemmati}, {Scarlata}, {Salvato}, {McCracken},
  {LeFevre}, \& {Scoville}}]{2012ApJ...760..128M}
{Mallery}, R.~P., {Mobasher}, B., {Capak}, P., {et~al.} 2012, \apj, 760, 128,
  \dodoi{10.1088/0004-637X/760/2/128}

\bibitem[{{Mason} {et~al.}(2018){Mason}, {Treu}, {de Barros}, {Dijkstra},
  {Fontana}, {Mesinger}, {Pentericci}, {Trenti}, \&
  {Vanzella}}]{2018ApJ...857L..11M}
{Mason}, C.~A., {Treu}, T., {de Barros}, S., {et~al.} 2018, \apjl, 857, L11,
  \dodoi{10.3847/2041-8213/aabbab}

\bibitem[{{Matsuda} {et~al.}(2004){Matsuda}, {Yamada}, {Hayashino}, {Tamura},
  {Yamauchi}, {Ajiki}, {Fujita}, {Murayama}, {Nagao}, {Ohta}, {Okamura},
  {Ouchi}, {Shimasaku}, {Shioya}, \& {Taniguchi}}]{Matsuda2004}
{Matsuda}, Y., {Yamada}, T., {Hayashino}, T., {et~al.} 2004, \aj, 128, 569,
  \dodoi{10.1086/422020}

\bibitem[{{Matsuda} {et~al.}(2012){Matsuda}, {Yamada}, {Hayashino}, {Yamauchi},
  {Nakamura}, {Morimoto}, {Ouchi}, {Ono}, {Umemura}, \& {Mori}}]{Matsuda2012}
---. 2012, \mnras, 425, 878, \dodoi{10.1111/j.1365-2966.2012.21143.x}

\bibitem[{{Matthee} {et~al.}(2015){Matthee}, {Sobral}, {Santos},
  {R{\"o}ttgering}, {Darvish}, \& {Mobasher}}]{2015MNRAS.451..400M}
{Matthee}, J., {Sobral}, D., {Santos}, S., {et~al.} 2015, \mnras, 451, 400,
  \dodoi{10.1093/mnras/stv947}

\bibitem[{{Neufeld}(1990)}]{1990ApJ...350..216N}
{Neufeld}, D.~A. 1990, \apj, 350, 216, \dodoi{10.1086/168375}

\bibitem[{{Ocvirk} {et~al.}(2016){Ocvirk}, {Gillet}, {Shapiro}, {Aubert},
  {Iliev}, {Teyssier}, {Yepes}, {Choi}, {Sullivan}, {Knebe}, {Gottl{\"o}ber},
  {D'Aloisio}, {Park}, {Hoffman}, \& {Stranex}}]{2016MNRAS.463.1462O}
{Ocvirk}, P., {Gillet}, N., {Shapiro}, P.~R., {et~al.} 2016, \mnras, 463, 1462,
  \dodoi{10.1093/mnras/stw2036}

\bibitem[{{Ocvirk} {et~al.}(2020){Ocvirk}, {Aubert}, {Sorce}, {Shapiro},
  {Deparis}, {Dawoodbhoy}, {Lewis}, {Teyssier}, {Yepes}, {Gottl{\"o}ber},
  {Ahn}, {Iliev}, \& {Hoffman}}]{2020MNRAS.496.4087O}
{Ocvirk}, P., {Aubert}, D., {Sorce}, J.~G., {et~al.} 2020, \mnras, 496, 4087,
  \dodoi{10.1093/mnras/staa1266}

\bibitem[{{Ono} {et~al.}(2012){Ono}, {Ouchi}, {Mobasher}, {Dickinson},
  {Penner}, {Shimasaku}, {Weiner}, {Kartaltepe}, {Nakajima}, {Nayyeri},
  {Stern}, {Kashikawa}, \& {Spinrad}}]{2012ApJ...744...83O}
{Ono}, Y., {Ouchi}, M., {Mobasher}, B., {et~al.} 2012, \apj, 744, 83,
  \dodoi{10.1088/0004-637X/744/2/83}

\bibitem[{{Ouchi} {et~al.}(2010){Ouchi}, {Shimasaku}, {Furusawa}, {Saito},
  {Yoshida}, {Akiyama}, {Ono}, {Yamada}, {Ota}, {Kashikawa}, {Iye}, {Kodama},
  {Okamura}, {Simpson}, \& {Yoshida}}]{Ouchi2010}
{Ouchi}, M., {Shimasaku}, K., {Furusawa}, H., {et~al.} 2010, \apj, 723, 869,
  \dodoi{10.1088/0004-637X/723/1/869}

\bibitem[{{Park} {et~al.}(2021){Park}, {Jung}, {Song}, {Ocvirk}, {Shapiro},
  {Dawoodbhoy}, {Iliev}, {Ahn}, {Bianco}, \& {Kim}}]{2021ApJ...922..263P}
{Park}, H., {Jung}, I., {Song}, H., {et~al.} 2021, \apj, 922, 263,
  \dodoi{10.3847/1538-4357/ac2f4b}

\bibitem[{Partridge \& Peebles(1967)}]{Partridge1967}
Partridge, R.~B., \& Peebles, P. J.~E. 1967, \apj, 147, 868,
  \dodoi{10.1086/149079}

\bibitem[{{Pentericci} {et~al.}(2011){Pentericci}, {Fontana}, {Vanzella},
  {Castellano}, {Grazian}, {Dijkstra}, {Boutsia}, {Cristiani}, {Dickinson},
  {Giallongo}, {Giavalisco}, {Maiolino}, {Moorwood}, {Paris}, \&
  {Santini}}]{Pentericci2011}
{Pentericci}, L., {Fontana}, A., {Vanzella}, E., {et~al.} 2011, \apj, 743, 132,
  \dodoi{10.1088/0004-637X/743/2/132}

\bibitem[{{Qin} {et~al.}(2022){Qin}, {Wyithe}, {Oesch}, {Illingworth},
  {Leonova}, {Mutch}, \& {Naidu}}]{2022MNRAS.510.3858Q}
{Qin}, Y., {Wyithe}, J. S.~B., {Oesch}, P.~A., {et~al.} 2022, \mnras, 510,
  3858, \dodoi{10.1093/mnras/stab3733}

\bibitem[{{Rosdahl} \& {Blaizot}(2012)}]{2012MNRAS.423..344R}
{Rosdahl}, J., \& {Blaizot}, J. 2012, \mnras, 423, 344,
  \dodoi{10.1111/j.1365-2966.2012.20883.x}

\bibitem[{{Sadoun} {et~al.}(2017){Sadoun}, {Zheng}, \&
  {Miralda-Escud{\'e}}}]{2017ApJ...839...44S}
{Sadoun}, R., {Zheng}, Z., \& {Miralda-Escud{\'e}}, J. 2017, \apj, 839, 44,
  \dodoi{10.3847/1538-4357/aa683b}

\bibitem[{{Schaerer} \& {Verhamme}(2008)}]{2008A&A...480..369S}
{Schaerer}, D., \& {Verhamme}, A. 2008, \aap, 480, 369,
  \dodoi{10.1051/0004-6361:20078913}

\bibitem[{{Semelin} {et~al.}(2007){Semelin}, {Combes}, \&
  {Baek}}]{2007A&A...474..365S}
{Semelin}, B., {Combes}, F., \& {Baek}, S. 2007, \aap, 474, 365,
  \dodoi{10.1051/0004-6361:20077965}

\bibitem[{{Seon} {et~al.}(2022){Seon}, {Song}, \&
  {Chang}}]{2022ApJS..259....3S}
{Seon}, K.-i., {Song}, H., \& {Chang}, S.-J. 2022, \apjs, 259, 3,
  \dodoi{10.3847/1538-4365/ac3af1}

\bibitem[{{Smith} {et~al.}(2022){Smith}, {Kannan}, {Garaldi}, {Vogelsberger},
  {Pakmor}, {Springel}, \& {Hernquist}}]{2022MNRAS.512.3243S}
{Smith}, A., {Kannan}, R., {Garaldi}, E., {et~al.} 2022, \mnras, 512, 3243,
  \dodoi{10.1093/mnras/stac713}

\bibitem[{{Smith} {et~al.}(2021){Smith}, {Kannan}, {Tacchella}, {Vogelsberger},
  {Hernquist}, {Marinacci}, {Sales}, {Torrey}, {Li}, {Yeh}, \&
  {Qi}}]{2021arXiv211113721S}
{Smith}, A., {Kannan}, R., {Tacchella}, S., {et~al.} 2021, arXiv e-prints,
  arXiv:2111.13721.
\newblock \doarXiv{2111.13721}

\bibitem[{{Stark} {et~al.}(2011){Stark}, {Ellis}, \&
  {Ouchi}}]{2011ApJ...728L...2S}
{Stark}, D.~P., {Ellis}, R.~S., \& {Ouchi}, M. 2011, \apjl, 728, L2,
  \dodoi{10.1088/2041-8205/728/1/L2}

\bibitem[{{Steidel} {et~al.}(2000){Steidel}, {Adelberger}, {Shapley},
  {Pettini}, {Dickinson}, \& {Giavalisco}}]{Steidel2000}
{Steidel}, C.~C., {Adelberger}, K.~L., {Shapley}, A.~E., {et~al.} 2000, \apj,
  532, 170, \dodoi{10.1086/308568}

\bibitem[{{Tasitsiomi}(2006)}]{2006ApJ...645..792T}
{Tasitsiomi}, A. 2006, \apj, 645, 792, \dodoi{10.1086/504460}

\bibitem[{{Tilvi} {et~al.}(2014){Tilvi}, {Papovich}, {Finkelstein}, {Long},
  {Song}, {Dickinson}, {Ferguson}, {Koekemoer}, {Giavalisco}, \&
  {Mobasher}}]{2014ApJ...794....5T}
{Tilvi}, V., {Papovich}, C., {Finkelstein}, S.~L., {et~al.} 2014, \apj, 794, 5,
  \dodoi{10.1088/0004-637X/794/1/5}

\bibitem[{{Tilvi} {et~al.}(2020){Tilvi}, {Malhotra}, {Rhoads}, {Coughlin},
  {Zheng}, {Finkelstein}, {Veilleux}, {Mobasher}, {Wang}, {Probst}, {Swaters},
  {Hibon}, {Joshi}, {Zabl}, {Jiang}, {Pharo}, \& {Yang}}]{2020ApJ...891L..10T}
{Tilvi}, V., {Malhotra}, S., {Rhoads}, J.~E., {et~al.} 2020, \apjl, 891, L10,
  \dodoi{10.3847/2041-8213/ab75ec}

\bibitem[{{Treu} {et~al.}(2013){Treu}, {Schmidt}, {Trenti}, {Bradley}, \&
  {Stiavelli}}]{2013ApJ...775L..29T}
{Treu}, T., {Schmidt}, K.~B., {Trenti}, M., {Bradley}, L.~D., \& {Stiavelli},
  M. 2013, \apjl, 775, L29, \dodoi{10.1088/2041-8205/775/1/L29}

\bibitem[{{Verhamme} {et~al.}(2008){Verhamme}, {Schaerer}, {Atek}, \&
  {Tapken}}]{2008A&A...491...89V}
{Verhamme}, A., {Schaerer}, D., {Atek}, H., \& {Tapken}, C. 2008, \aap, 491,
  89, \dodoi{10.1051/0004-6361:200809648}

\bibitem[{{Verhamme} {et~al.}(2006){Verhamme}, {Schaerer}, \&
  {Maselli}}]{2006A&A...460..397V}
{Verhamme}, A., {Schaerer}, D., \& {Maselli}, A. 2006, \aap, 460, 397,
  \dodoi{10.1051/0004-6361:20065554}

\bibitem[{{Visbal} \& {McQuinn}(2018)}]{2018ApJ...863L...6V}
{Visbal}, E., \& {McQuinn}, M. 2018, \apjl, 863, L6,
  \dodoi{10.3847/2041-8213/aad5e6}

\bibitem[{{Wisotzki} {et~al.}(2018){Wisotzki}, {Bacon}, {Brinchmann},
  {Cantalupo}, {Richter}, {Schaye}, {Schmidt}, {Urrutia}, {Weilbacher},
  {Akhlaghi}, {Bouch{\'e}}, {Contini}, {Guiderdoni}, {Herenz}, {Inami},
  {Kerutt}, {Leclercq}, {Marino}, {Maseda}, {Monreal-Ibero}, {Nanayakkara},
  {Richard}, {Saust}, {Steinmetz}, \& {Wendt}}]{2018Natur.562..229W}
{Wisotzki}, L., {Bacon}, R., {Brinchmann}, J., {et~al.} 2018, \nat, 562, 229,
  \dodoi{10.1038/s41586-018-0564-6}

\bibitem[{{Yajima} {et~al.}(2012{\natexlab{a}}){Yajima}, {Li}, {Zhu}, \&
  {Abel}}]{2012MNRAS.424..884Y}
{Yajima}, H., {Li}, Y., {Zhu}, Q., \& {Abel}, T. 2012{\natexlab{a}}, \mnras,
  424, 884, \dodoi{10.1111/j.1365-2966.2012.21228.x}

\bibitem[{{Yajima} {et~al.}(2012{\natexlab{b}}){Yajima}, {Li}, {Zhu}, \&
  {Abel}}]{Yajima2012}
---. 2012{\natexlab{b}}, \mnras, 424, 884,
  \dodoi{10.1111/j.1365-2966.2012.21228.x}

\bibitem[{{Yang} {et~al.}(2016){Yang}, {Malhotra}, {Gronke}, {Rhoads},
  {Dijkstra}, {Jaskot}, {Zheng}, \& {Wang}}]{2016ApJ...820..130Y}
{Yang}, H., {Malhotra}, S., {Gronke}, M., {et~al.} 2016, \apj, 820, 130,
  \dodoi{10.3847/0004-637X/820/2/130}

\bibitem[{{Yang} {et~al.}(2014){Yang}, {Zabludoff}, {Jahnke}, \&
  {Dav{\'e}}}]{2014ApJ...793..114Y}
{Yang}, Y., {Zabludoff}, A., {Jahnke}, K., \& {Dav{\'e}}, R. 2014, \apj, 793,
  114, \dodoi{10.1088/0004-637X/793/2/114}

\bibitem[{{Zheng} {et~al.}(2010){Zheng}, {Cen}, {Trac}, \&
  {Miralda-Escud{\'e}}}]{2010ApJ...716..574Z}
{Zheng}, Z., {Cen}, R., {Trac}, H., \& {Miralda-Escud{\'e}}, J. 2010, \apj,
  716, 574, \dodoi{10.1088/0004-637X/716/1/574}

\bibitem[{{Zheng} {et~al.}(2011){Zheng}, {Cen}, {Weinberg}, {Trac}, \&
  {Miralda-Escud{\'e}}}]{2011ApJ...739...62Z}
{Zheng}, Z., {Cen}, R., {Weinberg}, D., {Trac}, H., \& {Miralda-Escud{\'e}}, J.
  2011, \apj, 739, 62, \dodoi{10.1088/0004-637X/739/2/62}

\bibitem[{{Zheng} \& {Miralda-Escud{\'e}}(2002)}]{2002ApJ...578...33Z}
{Zheng}, Z., \& {Miralda-Escud{\'e}}, J. 2002, \apj, 578, 33,
  \dodoi{10.1086/342400}

\bibitem[{{Zheng} {et~al.}(2017){Zheng}, {Wang}, {Rhoads}, {Infante},
  {Malhotra}, {Hu}, {Walker}, {Jiang}, {Jiang}, {Hibon}, {Gonzalez}, {Kong},
  {Zheng}, {Galaz}, \& {Barrientos}}]{2017ApJ...842L..22Z}
{Zheng}, Z.-Y., {Wang}, J., {Rhoads}, J., {et~al.} 2017, \apjl, 842, L22,
  \dodoi{10.3847/2041-8213/aa794f}

\end{thebibliography}
\end{document}